\documentclass[aps]{revtex4}
\def\pochh #1#2{{(#1)\raise-4pt\hbox{$\scriptstyle#2$}}}
\def\binom#1#2{\left(\begin{array}{c}#1\\#2\end{array}\right)}
\font\mas=msbm10 \font\mass=msbm7 \textfont8=\mas \scriptfont8=\mass
\def\masy{\fam8}
\def\ms#1{{\masy#1}}
\usepackage{graphicx} 
\usepackage{dcolumn} 

\begin {document}

\title {The Functional Integral with Unconditional Wiener Measure  for Anharmonic Oscillator }
\author{J. Boh\' a\v cik}
\email{bohacik@savba.sk} \affiliation{Institute of Physics, Slovak
Academy of Sciences, D\' ubravsk\' a cesta 9, 845 11 Bratislava,
Slovakia.}
\author{P. Pre\v snajder}
\email{presnajder@fmph.uniba.sk} \affiliation{Department of
Theoretical Physics and Physics Education, Faculty of Mathematics,
Physics and Informatics, Comenius University, Mlynsk\' a dolina F2,
842 48 Bratislava, Slovakia.}

\begin{abstract}
In this article we propose the calculation of the unconditional
 Wiener measure functional integral with a term of the fourth order in
the exponent by an alternative method as in the conventional
perturbative approach. In contrast to the conventional perturbation
theory, we expand into  power series the term linear in the
integration variable in the exponent. In such a case we can profit
from the representation of the integral in question by the parabolic
cylinder functions. We show that in such a case the series
expansions are uniformly convergent and we find recurrence relations
for the Wiener functional integral in the $N$ - dimensional
approximation. In continuum limit we find that the generalized
Gelfand - Yaglom differential equation with solution yields the
desired functional integral (similarly as the standard Gelfand -
Yaglom differential equation yields the functional integral for
linear harmonic oscillator).

\end{abstract}
\maketitle

\section{Introduction}

We will define the continuum functional integral as the limit of a
finite dimensional integral. This finite dimensional integral is
derived from the continuum one by the time-slicing method. We avoid
so problems with continuum integral measure because we consider
continuum limit of the result of a finite dimensional integral. In
order to evaluate finite dimensional integrals we must first solve
the problem of the calculation of one dimensional integral:
\begin {equation}
I_1=\int\limits _{-\infty}^{+\infty}\;dx\;\exp\{-(A x^4+B x^2+C
x)\}\ ,
 \label {int1d}
\end {equation}
where $Re\: A>0$. The exact analytical result for this integral is
not known yet, therefore one uses approximative methods of
calculation.

\noindent The usual perturbative approach is based on Taylor's
decomposition of the fourth order term with consecutive replacements
of the order of integration and summation:
\begin {equation}
I_1=\sum\limits _{n=0} ^{\infty} \frac{(-A)^n} {n!}\int\limits
_{-\infty}^{+\infty}\;dx\;x^{4 n}\exp\{-(B x^2+C x)\}
\end {equation}
\noindent The above integrals can be calculated, but their sum is
divergent.

 However, $I_1=I_1(A,B,C)$ is an entire function for any complex values
 of $B$ and $C$, since there exist all integrals
 \[ \partial_C^n\partial_B^m I_1(A,B,C)=(-1)^{n+m}
 \int\limits _{-\infty}^{+\infty}
 \;dx\;x^{2m+n}\exp\{-(A x^4+Bx^2+Cx)\}\ ,\ Re A > 0.
 \]
Consequently, the power series expansions of $I_1=I_1(A,B,C)$ in $C$
and/or $B$ have an infinite radius of convergence (and in particular
they are uniformly convergent on any compact set of values of $C$
and/or $B$). We shall frequently use the power series expansion in
$C$:
\begin {equation}
I_1=\sum\limits _{n=0} ^{\infty} \frac{(-C)^n} {n!}\int\limits
_{-\infty}^{+\infty}\;dx\;x^n\exp\{-(A x^4+B x^2)\}
 \label{sim1}
\end {equation}

The similar idea of the first order term expansions in the fourth
order action  was used by Tuszy\' nski et al.\cite{tusz} for an
evaluations of a non-Gaussian models for critical fluctuations of
the Landau - Ginsburg model of phase transitions.

The integral in (\ref{sim1}) can be expressed in terms of the
parabolic cylinder function $D_{\nu}(z)$,  (see, for instance, \cite
{prud}). For $n$ odd, due to symmetry of the integrand the integrals
are zero, for $n$ even, $n=2m$ we have:
 \[\frac{e^{z^2/4}}{(\sqrt{2A})^{m+1/2}}\ \Gamma(m+1/2)\ D_{-m-1/2}(z)\ =\
\int_0^\infty\;dy\;
 y^{m-1/2}\exp\{-Ay^2-By\}\ , \]
where$$z=\frac{B}{\sqrt{2A}}\ .$$ \noindent Explicitly, for the
Eq.(\ref{sim1}) we have:
 \begin {equation}
 I_1=e^{z^2/4}\ \frac{\Gamma(1/2)}{(2A)^{1/4}}\sum\limits _{m=0} ^{\infty}
 \frac{(\xi)^m}{m!}D_{-m-1/2}(z)
 \label{s2}\ ,\ \xi=\frac{C^2}{4\sqrt{2A}}\ .
\end {equation}
This sum is convergent for any values of $C$, $B$ and $A$ positive.

The convergence of the infinite series in Eq. (\ref{s2}) can be
shown as follows.
 For $|z|$ finite, $|z|<\sqrt{|\nu|}$ and
$\mid arg(-\nu)\mid \leq \pi/2$ and if $\mid \nu \mid \rightarrow
\infty$, the following asymptotic relation is valid \cite{bateman}:
\begin {equation}
D_{\nu}(z)=\frac{1}{\sqrt{2}}\; \exp\left[\frac{\nu}{2}
(\ln{(-\nu)}-1) -\sqrt{-\nu}\;
z\right]\left[1+O\left(\frac{1}{\sqrt{\mid \nu\mid }}\right)\right]\
. \label {ass1}
\end {equation}
The $m$ term  of the sum in Eq. (\ref {s2}) possesses the
asymptotic:
\begin {equation}
\frac{1}{m!}\exp\left[-\frac{(m+1/2)}{2}(\ln{(m+1/2)}-1)
-\sqrt{(m+1/2)}\;z+m\ln\xi \right]\ .
\end {equation}

 This means, applying the
Bolzano-Cauchy criterium that the sum in Eq. (\ref {s2}) is not only
absolutely, but uniformly convergent for the finite values of the
constants of the integral (\ref {int1d}).

\section{Evaluation of the functional integral by time slicing method}

We suppose that Gaussian integration over momenta is done. Our aim
is to evaluate the continuum \emph{unconditional Wiener measure}
functional integral:
$$\mathcal{Z} = \int [\mathcal{D}\varphi(x)]\exp (-\mathcal{S})\ ,$$
where the continuum action contains the fourth-order term:
\begin {equation}
\mathcal{S} =\int \limits _0^\beta d\tau \left[c/2
\left(\frac{\partial\varphi(\tau)}{\partial\tau}\right)^2+b\varphi(\tau)^2
+a\varphi(\tau)^4\right]\ . \label{acti}
\end {equation}

The functional integral $\mathcal{Z}$ is defined by a limiting
procedure from the finite dimensional integral $\mathcal{Z}_{N}$,
obtained from the continuum integral, when the infinite measure
$[\mathcal{D}\varphi(x)]$ is replaced by the finite dimensional
measure $\prod_{i=1}^N\ d \varphi_i(x)$ \cite{dem}:

\begin {equation}
\mathcal{Z}_{N}=\int\limits _{-\infty}^{+\infty} \prod \limits
_{i=1}^N\left(\frac{d\varphi_i}{\sqrt{\frac{2\pi\triangle}{c}}}\right)
\exp\left\{-\sum\limits _{i=1}^N \triangle\left[c/2
\left(\frac{\varphi_i-\varphi_{i-1}}{\triangle}\right)^2
+b\varphi_i^2+a\varphi_i^4\right]\right\} , \label{findim}
\end {equation}

\noindent
 where $\triangle=\beta/N$.
 The \emph{unconditional measure} integration is characterized by integration over
 variable $\varphi_N$. This is the only difference from conditional measure
 integration, when the $\varphi_N$ variable is fixed.
The detailed discussion of the conditional Wiener measure case is
technically more involved and was not in the program of this
article. We evaluated this case in a rather simplified form in the
Appendix 4 of our article I \cite{paper1}. We included a brief
discussion of this case in conclusions. The continuum unconditional
Wiener measure functional integral is defined by the formal limit:
$$\mathcal{Z} = \lim_{N\rightarrow \infty}\;\mathcal{Z}_{N}\ .$$

To evaluate this limit we follow the idea of the Gelfand-Yaglom
proof of the functional integral for the harmonic oscillator
\cite{bkz}, based on the iterative procedure for the finite
dimensional representation of the functional integral. The idea of
$N$ dimensional integration of (\ref{findim}) is explained in
Appendix A, we  quote here the result:
\begin {equation}
\mathcal{Z}_{N} =
\left[2\pi(1+b\triangle^2/c)\right]^{-\frac{N-1}{2}}
\left[2\pi(1/2+b\triangle^2/c)\right]^{-1/2}
\sum\limits_{k_1,\cdots,k_{N-1}=0}^\infty \prod \limits _{i=1}^N \;
\left[ \frac{\left(\xi_i\right)^{2k_{i}}}{(2k_{i})!}
\Gamma(k_{i-1}+k_{i}+1/2)\mathcal{D}_{-k_{i-1}-k_{i}-1/2}\;(z_i)\right]\
, \label{fin1}
\end {equation}

 \noindent where $k_0 \equiv k_N \equiv 0,$
$\xi_1 = \xi_2 = \cdots = \xi_{N-2} = \xi=(1+b\triangle^2/c)^{-1},$
$\xi_{N-1}=(1+b\triangle^2/c)^{-1/2}(1/2+b\triangle^2/c)^{-1/2},$
 $\xi_{N} =0\ ,$

\noindent $z_1 = z_2 = \cdots = z_{N-1} =
z=c(1+b\triangle^2/c)/\sqrt{2a\triangle^3},$
$z_N=c(1/2+b\triangle^2/c)/\sqrt{2a\triangle^3}.$

\section{generalized Gelfand -- Yaglom equation (GGYE)}

Let us rewrite the result for the $N$ dimensional integral
(\ref{fin1}) in the form:
\begin {equation}
\mathcal{Z}_{N} = \left[\prod_{i=0}^{N}2(1+b\triangle^2/c)
\omega_i\right]^{-\frac{1}{2}} \; \mathcal{S}_{N} \label{ndim}
\end {equation}
with
\begin {equation}
\mathcal{S}_{N} = \sum\limits_{k_1,\cdots,k_{N-1}=0}^\infty \prod \limits _{i=1}^N \;
\left[ \frac{\left(\xi_i\right)^{2k_{i}}}{(2k_{i})!}\ \Gamma^{-1}(1/2) \sqrt{\omega_i}\
\Gamma(k_{i-1}+k_{i}+1/2)\mathcal{D}_{-k_{i-1}-k_{i}-1/2}\;(z_i)\right]\
, \label{ndim1}
\end {equation}

\noindent where the constants and symbols in the above  relation are
connected to the constants of the model by the relations:
$\omega_i=1-A^2/\omega_{i-1}$,
$\omega_0=(1/2+b\triangle^2/c)/(1+b\triangle^2/c)$,
$A=\frac{1}{2(1+b\triangle^2/c)}$. We prove the above form
(\ref{ndim}) of the $N$ dimensional integral (\ref{fin1}) later, now
we use  Eq. (\ref{ndim}) for explanation of the Gelfand -- Yaglom
procedure of the construction of the difference equation. This
difference equation is converted to differential equation in the
continuum limit $N\rightarrow \infty\ .$

Let us define functions $F_k$ by:

\begin {equation}
F_k =
\frac{\prod\limits^{k}_{i=0}2(1+b\triangle^2/c)\omega_i}{\mathcal{S}^2_{k}}
\label{gjf}
\end {equation}
The function $F_{k}$ is defined from the relation of the $N$
dimensional integral (\ref{ndim}) with $\triangle$ fixed and the
variable $N\rightarrow k$. The quantity $\mathcal{Z}_N$ given in Eq.
(\ref{fin1}) is related to $F_N$ as follows:

$$\mathcal{Z}_N = \frac{1}{\sqrt{F_N}}$$
The aim of the Gelfand-Yaglom construction is to find the continuum
limit of the difference equation for the function $F_k.$ Solution of
this differential equation is connected to the continuum functional
integral by:
$$\mathcal{Z(\beta)} =\frac{1}{\sqrt{F(\beta)}},$$
where $\beta$ is the upper bound of the time interval in the action
(\ref{acti}).

The idea of the GGYE construction is based on the recurrence form for
the factor $\omega_i\ .$ We replace $\omega_i$ by
 the functions $F_{k\pm 1}$ and $\mathcal{S}_{k, k\pm 1}\ .$
Pedagogical descriptions of this procedure can be found in Appendix
B, there ia a proof of the lemma:

\noindent
\textit{Lemma.} Let $F_k$ be the function defined by:
\begin {equation}
F_k =
\frac{\prod\limits^{k}_{i=0}2(1+b\triangle^2/c)\omega_i}{\mathcal{S}^2_{k}}
\end {equation}
with $\omega_i$ defined by recurrence  relation: $$\omega_i =
1-A^2/\omega_{i-1}\ ,$$ and
$\omega_0=(1/2+b\triangle^2/c)/(1+b\triangle^2/c)$,
$A=\frac{1}{2(1+b\triangle^2/c)}.$ The constants $b, c, \triangle$
are parameters of the model.

\noindent Let in continuum limit the following condition is valid:
\begin {equation}
\lim_{\triangle \rightarrow 0}(\triangle \ \mathcal{O}_1+\triangle^2
\ \mathcal{O}_2) = 0\ ,
\end {equation}
\noindent where
\begin {eqnarray}
\mathcal{O}_1 &=&-
\frac{4b}{c}F_k\left(\frac{\mathcal{S}_{k+1}-\mathcal{S}_k}{\triangle
\mathcal{S}_{k+1}}\right)
+2\frac{(F_k-F_{k-1})}{\triangle}\left[\frac{b}{c}+\left(\frac{\mathcal{S}_{k+1}-\mathcal{S}_k}{\triangle
\mathcal{S}_{k+1}}\right)^2\right]
+4F_{k-1}\left(\frac{\mathcal{S}_{k+1}-\mathcal{S}_k}{\triangle
\mathcal{S}{k+1}}\right)\left(\frac{\mathcal{S}_{k+1}-2\mathcal{S}_k+\mathcal{S}_{k-1}}{\triangle^2
\mathcal{S}_{k+1}}\right) \nonumber
\\ \noalign{\vskip8pt}
\mathcal{O}_2 &=&
\frac{2b}{c}F_k\left(\frac{\mathcal{S}_{k+1}-\mathcal{S}_k}{\triangle
\mathcal{S}_{k+1}}\right)^2-
F_{k-1}\left(\frac{\mathcal{S}_{k+1}-2\mathcal{S}_k+\mathcal{S}_{k-1}}{\triangle^2
\mathcal{S}_{k+1}}\right)^2.
\end {eqnarray}
\noindent Let in the limit $\triangle\rightarrow 0,$ the  functions
$F(\tau)$ and $\mathcal{S}(\tau)$ are the continuum limit of the
function $F_k$ and $\mathcal{S}_k$, when we define: $\lim_{\triangle
\rightarrow 0}k.\triangle = \tau\ .$

Then $F(\tau)$ is the solution of the differential
equation:
\begin {equation}
\frac{\partial^2}{\partial
\tau^2}F(\tau)+4\left(\frac{\partial}{\partial \tau}F(\tau)\right)
\; \left(\frac{\partial}{\partial
\tau}\ln{\mathcal{S}(\tau)}\right)=
F(\tau)\left(\frac{2b}{c}-2\frac{\partial^2}{\partial
\tau^2}\ln{\mathcal{S}(\tau)}-4(\frac{\partial}{\partial
\tau}\ln{\mathcal{S}(\tau)})^2\right)\ ,\ \tau\in (0, \beta)
\label{gy1}
\end {equation}
with initial conditions
\begin {eqnarray}
F(0) &=& \frac{1}{\mathcal{S}^2(0)},
\\
\frac{\partial}{\partial \tau}F(\tau)\Big |_{\tau=0} &=&
-\frac{\partial}{\partial \tau}\left(\frac{1}{\mathcal{S}^2(\tau)}\right)\Big
|_{\tau=0}\ .\nonumber
\end{eqnarray}

\noindent
The nontrivial dynamics is hidden in the function $\mathcal{S}(\tau)$.

\textit{Note:}When $\mathcal{S}(\tau)$ is known exactly the above equation can be simplified by the substitution:
$$F(\tau) = \frac{y(\tau)}{\mathcal{S}^2(\tau)}\ .$$
For the new variable $y(\tau)$ we find a simple equation:
\begin {equation}
\frac{\partial^2}{\partial \tau^2}y(\tau) =
y(\tau)\left(\frac{2b}{c}  \right)\ , \label{gy2}
\end {equation}
accompanied  by  initial conditions:

$$\left. y(0) = 1 ,\
\frac{\partial y(\tau)}{\partial \tau}\right|_{\tau=0}
= 0.
$$

\noindent Thus, in the case when function $\mathcal{S}(\tau)$ is
known exactly, the problem of the functional integral calculation is
trivial.

Problems arise in situations, when $\mathcal{S}(\tau)$ is known
approximately, as a result of a perturbative approach.
Below we define a reasonable approximation of $\mathcal{S}(\tau)$
valid in the proximity of $\tau=0.$ But for finite (large)
$\tau=\beta$ this asymptotic expansion of $\mathcal{S}(\tau)$ does
not have to be valid. However, in our case, the development of the
function $F(\tau)$ from $\tau=0$ to $\tau=\beta$ is controlled by a
differential equation. Approximative knowledge of the function
$\mathcal{S}(\tau)$ leads to a more reliable result for $F(\beta)$
as the solution of Eq. (\ref{gy2}). This philosophy of the
calculation corresponds to ideas of evaluation of physical
quantities be the renormalization group approach.

\section{Evaluation of the function $\mathcal{S}_N$}

The exact result of $N-$ dimensional integration (\ref{fin1})
contains summations of products of parabolic cylinder functions. We
were not able to find a simple formula fur such a summation
(although, parabolic cylinder functions belong to the representation
of the group of the upper-triangle matrices, therefore, due to the
group theoretical background we believe to the simplification of the
product of two such functions). In what follows, we explain our
approach to provide the summation over indexes $k_i$. For a single
index $k_i$ we are dealing with the sum of the series:

\begin {equation}
\sum_{k_i=0}^{\infty}a_{k_i}\ ,\
a_{k_i}=\frac{\left(\xi\right)^{2k_{i}}}{(2k_{i})!}
\Gamma(k_{i-1}+k_{i}+1/2)\mathcal{D}_{-k_{i-1}-k_{i}-1/2}\;(z)
\Gamma(k_{i}+k_{i+1}+1/2)\mathcal{D}_{-k_{i}-k_{i+1}-1/2}\;(z)
\label{ass}
\end {equation}
The series (\ref{ass}) is uniformly convergent. For asymptotic
values of $k_i$ following the Stirling formula for logarithm of
gamma functions and the asymptotic relation (\ref{ass1}) for
parabolic cylinder functions we find \cite{paper1}:

\begin {eqnarray}
\ln a_{k_i}&=& -(k_i - \frac{k_{i-1}+k_{i+1}-1}{2})\ln k_i +
2k_i(\ln(\xi z) -\ln2 +1/2)-2\sqrt{k_i}\; z + (k_{i-1}+k_{i+1}+1)\ln z
\nonumber \\
&+& 1/2\ln{\pi} + z^2/2 + o(k_i^{-1/2})
\end {eqnarray}
The leading term of the above relation is $$\ln(a_{k_i})\sim -k_i
\ln(k_i), $$ and $$a_{k_i}\sim \frac{(\xi z)^{2k_i}}{k_i !}
\frac{(\sqrt{k_i}\; z)^{k_{i-1}+k_{i+1}}}{\exp{(2\sqrt{k_i}\; z)}}\
.$$ This asymptotic behavior of $a_{k_i}$ is sufficient for a proof
of the uniform convergence of the series. The convergence criteria
are fulfilled for an arbitrary group of sums over $k_i$ index in the
result (\ref{fin1}) for $N-$ dimensional integral.

In fact in (\ref{fin1}) we have an $N$-tuple sum of two index
quantities (closely related to $a_{k_i}$). Let us for simplicity
discuss the $N$ tuple sum of the two index quantities:
\begin {eqnarray}
& &\sum^{\infty}_{k_1,\cdots ,k_N=0}\alpha_{k_0,k_1}\alpha_{k_1,
k_2}\alpha_{k_2, k_3}\cdots \alpha_{k_{N-1}, k_N}\alpha_{k_N,
k_{N+1}},\nonumber \\
\alpha_{k_{i-1},k_{i}}&=&
\frac{\left(\xi_i\right)^{2k_{i}}}{(2k_{i})!}
\Gamma(k_{i-1}+k_{i}+1/2)\mathcal{D}_{-k_{i-1}-k_{i}-1/2}\;(z_i)
 \label{principal}
\end {eqnarray}
If each individual sum over $k_i$ exists and is finite, we divide
the sum over the index $k_i$ to a finite principal sum and a
remainder $\varepsilon_i(K_0)$, which can be made as small as
possible by suitable selection of the upper summation limit $K_0$,
comparing it to the principal part of the sum:
\begin {equation}
\sum^{\infty}_{k_i=0}\alpha_{k_{i-1}, k_i}\alpha_{k_i, k_{i+1}}=
\sum^{K_0}_{k_i=0}\alpha_{k_{i-1}, k_i}\alpha_{k_i, k_{i+1}}+
\sum^{\infty}_{k_i=K_0+1}\alpha_{k_{i-1}, k_i}\alpha_{k_i, k_{i+1}}=
(1+\varepsilon_i(K_0))\sum^{K_0}_{k_i=0}\alpha_{k_{i-1},
k_i}\alpha_{k_i, k_{i+1}}
\end {equation}
Let us define the " principal sum" $\Sigma(m,N-m)$ of
(\ref{principal}) as follows:  the first $m$ summations run to
infinity and the last $N-m$ summations run to $K_0$:
\begin {equation}
\Sigma(m,N-m)=\sum^{\infty}_{k_1,\cdots ,k_m=0}a_{k_1, k_2}a_{k_2,
k_3} \cdots a_{k_{m-1}, k_m}b_{k_m},
\end {equation}
where
\begin {equation}
b_{k_m} = \sum^{K_0}_{k_{m+1},\cdots ,k_N=0}a_{k_{m},
k_{m+1}}a_{k_{m+1}, k_{m+2}} \cdots a_{k_{N}, k_{N+1}}
\end {equation}
In this notation, the sum (\ref{principal}) is $\Sigma(N,0)$, our
aim is to estimate this sum by its principal sum $\Sigma(0,N)$, when
all summations run over finite range. Performing in $\Sigma(m,N-m)$
the sum over $k_m$, we obtain:
$$\Sigma(m,N-m) = (1+\varepsilon_m)\Sigma(m-1,N-m+1)\ .$$
 We have the inequality:
$$\Sigma(0,N)\leq \Sigma(N,0)\leq (1+\varepsilon)^N\Sigma(0,N)\ .$$
For any fixed $N$ we choose $K_0$ so that
$\varepsilon=\max{\varepsilon_i(K_0)}.$

\noindent Following this discussion, the finite dimensional integral
will converge if the upper bound $\varepsilon$ to remainders
approaches to zero as
$$\varepsilon \sim N^{-1-\theta},\ \theta>0.$$
This will guarantee the convergence of $\Sigma(0,N)$ to the
continuum integral.

 In what follows, we describe the idea of the evaluation of
"principal sum" for Eq. (\ref{fin1}) and how to estimate the
remainder to this leading term. The procedure of the summation
consists of the following steps. At first, we represent one of the
parabolic cylinder functions $\mathcal{D}_{-m-1/2}(z)$ by Poincar\'
e - type expansion \cite{bateman}, valid for real index and positive
argument of the function, under the assumption that the index of the
function is finite and the argument is going to infinity. For the
dimension of the integral sufficiently great this is consistent,
because $z \sim N^{3/2}.$ We have:

\begin {equation}
\mathcal{D}_{-m-1/2}(z)\;=\;
e^{z^2/4}\;z^{m+1/2}\;D_{-m-1/2}(z)\;=\;
\sum\limits_{j=0}^{\mathcal{J}}\; (-1)^j \;\frac{(m +
1/2)_{2j}}{j!\;(2z^2)^j} + \epsilon_{\mathcal{J}}(m,z)\ ,
\label{asex1}
\end {equation}
where $\epsilon_{\mathcal{J}}(m,z)$ is the remainder of the
Poincar\' e - type expansion of the $\mathcal{D}$ function. For
Poincar\' e - type expansion the upper bound of remainder was
calculated by Olver \cite{olver}. We use the improved upper bound
evaluated by Temme \cite{temme2}. The upper bound for remainder in
definition (\ref{asex1}) reads:
\begin {equation}
\mid \epsilon_{\mathcal{J}}(m,z)\mid\ \leq
\frac{2z^2}{z^2-2m}\frac{\pochh{m+1/2}{2\mathcal{J}}}{(\mathcal{J}-1)!\
(2z^2)^{\mathcal{J}}}\ _1F_2(\frac{\mathcal{J}}{2}, \frac{1}{2};
\frac{\mathcal{J}}{2}+1;
1-\frac{m^2}{z^2})\exp{\left(\frac{4\theta}{z^2-2m}\
_1F_2(\frac{1}{2},\frac{1}{2};\frac{3}{2};1-\frac{m^2}{z^2})\right)},
\label{temrem}
\end {equation}
where
$$\theta=\Big|\frac{m^2}{4}+\frac{3}{16}\Big|+
\frac{2m}{z^2}\left(1+\frac{m}{2z^2}\right)\frac{z^2}{(z^2-2m)^2}.$$

The estimate (\ref{temrem}) is valid for $2\sqrt{m}\leq z$
\cite{temme2}. Before insertion of the Poincar\' e - type expansion
(\ref{asex1}) into (\ref{ass}) for one of the $\mathcal{D}$,  we
divide the sum over $k_i$ into two parts. One, over finite $k_i$,
where Poincar\'e - type expansion is correct and the second, the
remainder,  small compared to the first part due to uniform
convergence:
\begin {eqnarray}
&&\sum\limits_{j=0}^{\mathcal{J}}\; \frac{(-1)^j }{j!\;(2z^2)^j}\
\sum_{k_i=0}^{K_0}\ \frac{\left(\xi\right)^{2k_{i}}}{(2k_{i})!}
\Gamma(k_{i-1}+k_{i}+1/2) \Gamma(k_{i}+k_{i+1}+2j+1/2)
\mathcal{D}_{-k_{i-1}-k_{i}-1/2}\;(z)
  +\\
 &+&\sum_{k_i=0}^{K_0}\ \frac{\left(\xi\right)^{2k_{i}}}{(2k_{i})!}
\Gamma(k_{i-1}+k_{i}+1/2) \Gamma(k_{i}+k_{i+1}+1/2)
\mathcal{D}_{-k_{i-1}-k_{i}-1/2}\;(z) \
\epsilon_{\mathcal{J}}(k_{i}+k_{i+1},z) + \nonumber \\
&+&\sum_{k_i=K_0+1}^{\infty}\
\frac{\left(\xi\right)^{2k_{i}}}{(2k_{i})!}
\Gamma(k_{i-1}+k_{i}+1/2) \Gamma(k_{i}+k_{i+1}+1/2)
\mathcal{D}_{-k_{i-1}-k_{i}-1/2}\;(z)\mathcal{D}_{-k_{i}-k_{i+1}-1/2}\;(z),\nonumber
\end {eqnarray}
where $2\sqrt{K_0} <z $. In the next steep we extend the summation
in the first, ("leading") term up to infinity by adding and
subtracting the terms, allowing to  sum over the index $k_i$
according to the relation\cite{bateman}:

\begin {equation}
e^{x^2/4}\sum\limits_{k=0}^{\infty}\;
\frac{\pochh{\nu}k}{k!}\;t^k\;D_{-\nu-k}(x)\;=\;e^{(x-t)^2/4}\;D_{-\nu}\;(x-t)\
. \label{dsum1}
\end {equation}
We summed up the product of the two functions $\mathcal{D}\ ,$ the
result is the function $\mathcal{D}$ with a new argument. The
pedagogical description of this procedure can be found in article I
\cite{paper1}. We show the idea of the evaluation of the leading
term in Appendix C.

We have shown, that the leading term for the $N$ dimensional
integral is of the form:
\begin {equation}
\mathcal{Z}_{N-1}^{cut}=
\left\{\prod\limits_{i=0}^{N-1}\frac{1}{\sqrt{2(1+b\triangle^2/c)\omega_i}}\right\}
\sum\limits_{\mu=0}^{\mathcal{J}}\frac{(-1)^{\mu}}{\mu!(2z^2)^{\mu}}\left(N\right)_0^{2\mu}
\label{rov12}
\end {equation}
where the symbol $\left(N\right)_0^{2\mu}$ is defined by the recurrence relation:
\begin {eqnarray}
&&(A Q_{\Lambda} Q_{\Lambda-1})^{2\mu-p}
\left(\Lambda\right)_{2\mu-p}^{2\mu} = \nonumber
\\ \noalign{\bigskip}
&=&\sum\limits_{j=0}^p\;\frac{a^{2\mu-j}_{2\mu-p}}{(A Q_{\Lambda}
Q_{\Lambda-1})^{p-j}}\sum\limits^{\mu}_{\lambda=[\frac{j+1}{2}]}
\left(A Q_{\Lambda-2}Q_{\Lambda-1}\right)^{2\lambda-j}\;
\left(\Lambda-1\right)_{2\lambda-j}^{2\lambda}\;
\binom{\mu}{\lambda}(Q_{\Lambda-1}^4)^{\mu-\lambda}\ ,\ p\in <0,
2\mu>
\label{recu}
\end {eqnarray}
where  the recurrence in $\left(\Lambda\right)_{2\mu-p}^{2\mu}$
starts from:
$$\left(1\right)_i^{2j} = \frac{1}{\omega_0^{2j}}a_i^{2j}.$$

Putting $\Lambda = N,$ and $p = 2\mu$ in Eq.(\ref{recu}) we obtain
the relation for the leading part of the $N-$ dimensional integral,
which in the continuum limit coincides with the functional integral
searched, if the remainder of the leading part in the continuum
limit disappears. The sum over index $\mu$ in Eq.(\ref{rov12}) is an
asymptotic sum, therefore the upper summation limit must be taken
symbolically. The estimate of the remainder to
$\mathcal{Z}_{N-1}^{cut}$ is discussed in Appendix D. We have shown,
that this remainder can be made smaller than $N^{-1-\theta},\ \theta
> 0$, which is the necessary condition for the vanishing remainder in the
continuum limit $N\rightarrow \infty$.

\section{Solution of the recurrence relation for $\left(\Lambda\right)_{2\mu-p}^{2\mu}$}

To evaluate the $N$ dimensional integral, we solve the recurrence
relation (\ref{recu}) for arbitrary value $\mu=d$:
\begin {eqnarray}
&&(A Q_{\Lambda} Q_{\Lambda-1})^{2d-p}
\left(\Lambda\right)_{2d-p}^{2d} = \nonumber
\\ \noalign{\bigskip}
&=&\sum\limits_{j=0}^p\;\frac{a^{2d-j}_{2d-p}}{(A Q_{\Lambda}
Q_{\Lambda-1})^{p-j}}\sum\limits^{d}_{\lambda=[\frac{j+1}{2}]}
\left(A Q_{\Lambda-2}Q_{\Lambda-1}\right)^{2\lambda-j}\;
\left(\Lambda-1\right)_{2\lambda-j}^{2\lambda}\;
\binom{d}{\lambda}(Q_{\Lambda-1}^4)^{d-\lambda}\ ,\ p\in <0,
2d> \nonumber
\end {eqnarray}
The right hand side of the equation is $(2d,\ p)$-th matrix element
of the product of three matrices. For fixed $d$ on the left hand
side of the equation, we read only the $d$-th column of a matrix,
which is recurrently tied to the matrix in the center of the product
on the left hand side. We define an auxiliary matrix $\ms
X^d(\Lambda)$ by the following matrix equation:

$$\ms X^d_{p,\mu}(\Lambda) = \sum\limits_{j=0}^p
\sum\limits^{\mu}_{\lambda=[\frac{j+1}{2}]}\; \ms
A^d_{p,j}(\Lambda-1)\ms C^d_{j,\lambda}(\Lambda-1)\ms
M^d_{\lambda,\mu}(\Lambda-1) .$$

To evaluate the matrix $\ms C^{\mu}(\Lambda)$, we must calculate
$\ms X^d(\Lambda)$ for all dimensions $d$ up to $\mu$, for each
dimension to extract $d$-th column of matrix $\ms X^d(\Lambda)$ and
 to compose from these columns the matrix $\ms C^{\mu}(\Lambda)$. We define such
linear operation as follows:

1. Let $\ms A^d$ and $\ms M^d$ are the matrices of the dimensions
$(2\mu+1)(2\mu+1)$ and $(\mu+1)(\mu+1)$ respectively,  $\ms C^d$ is
the matrix of dimensions $(2\mu+1)(\mu+1)$. These matrices possess
nonzero main minors of the dimensions $(2d+1)(2d+1)$, $(d+1)(d+1)$,
and $(2d+1)(d+1),$ respectively. The definition of the matrices $\ms
A^d(\Lambda)$ and $\ms M^d(\Lambda)$ is in Appendix E.

2. Matrix $\tilde{\ms X}^d$ is the one column matrix defined by the
relation:
$$\tilde{\ms X}^d(\Lambda) = \ms X^d(\Lambda)*\ms P^d,$$
where $\ms P^d$ is the projector of the d-th column of the matrix
$\ms X^d(\Lambda)$ into d-th column of the matrix $\tilde{\ms
X}^d(\Lambda)$. $\ms P^d$ is a matrix with a single nonzero term
$$\left\{\ms P^d\right\}_{d,k} = \delta_{d,k}.$$

3. The matrix $\ms X^d(\Lambda)$ is defined by relation:
$$\ms X^d(\Lambda) = \ms A^d(\Lambda-1)*\ms C^d(\Lambda-1)*\ms M^d(\Lambda-1).$$

4. Then, for $\ms C^d(\Lambda)$ we have the result:
$$\ms C^d(\Lambda) = \sum_{i_{\Lambda}=0}^d
\ms A^{d-i_{\Lambda}}(\Lambda-1)*\ms
C^{d-i_{\Lambda}}(\Lambda-1)*\tilde{\ms
M}^{d-i_{\Lambda}}(\Lambda-1).$$

5. After evaluation of the full recurrence we find:
\begin {eqnarray}
&&\ms C^d(\Lambda) = \sum_{i_{\Lambda}=0}^d
\sum_{i_{\Lambda-1}=0}^{d-i_{\Lambda}}\cdots
\sum_{i_2=0}^{d-i_{\Lambda}-i_{\Lambda-1}- \cdots -i_3}
\label{matr1}\\
\noalign{\bigskip}\nonumber &&\left\{\ms
A^{d-i_{\Lambda}}(\Lambda-1)*\ms
A^{d-i_{\Lambda}-i_{\Lambda-1}}(\Lambda-2)* \cdots
*\ms A^{d-i_{\Lambda}-i_{\Lambda-1}-\cdots-i_2}(1)\right\}*\ms C^{d-i_{\Lambda}-i_{\Lambda-1}-\cdots-i_2}(1)*
\\ \noalign{\bigskip}
\nonumber &&\left\{\tilde{\ms
M}^{d-i_{\Lambda}-i_{\Lambda-1}-\cdots-i_2}(1)*\cdots* \tilde{\ms
M}^{d-i_{\Lambda}-i_{\Lambda-1}}(\Lambda-2)* \tilde{\ms
M}^{d-i_{\Lambda}}(\Lambda-1)\right\}
\end {eqnarray}

\noindent The evaluations of multiple products of the matrices is
given in Appendix E. Remember that for the function
$\mathcal{S}_{\Lambda}$ defined in Eq. (\ref{rov12}) the only
important matrix element is $\left\{\ms
C(\Lambda)^{2\mu}\right\}_{2\mu, 2\mu}$  then we find the
result\cite{paper2}:

\begin {eqnarray}
&&\left\{\ms C(\Lambda)^{2\mu}\right\}_{2\mu, 2\mu} =
\sum_{i_{\Lambda}=0}^{\mu}\;
\sum_{i_{\Lambda-1}=0}^{\mu-i_{\Lambda}}\cdots
\sum_{i_2=0}^{\mu-i_{\Lambda}-i_{\Lambda-1}- \cdots -i_3}
\binom{I_3}{I_2} \cdots \binom{I_{\Lambda}}{I_{\Lambda-1}}
Q_2^{4(I_3-I_2)}\cdots
Q_{\Lambda-1}^{4(I_{\Lambda}-I_{\Lambda-1})}\times\nonumber
\\ \noalign{\bigskip}\nonumber
&&\times\sum_{j=0}^{2I_2}\sum_{\lambda =
\mid\frac{j+1}{2}\mid}^{I_2}
\frac{(4I_2-2j)!}{(2\mu-j)!}\frac{2^{4(i_2  + \cdots +
i_{\Lambda})}}{2^{(4\mu-2j)}}
\left\{\prod_{m=2}^{\Lambda}D^{i_m}_{\xi_m}
\left[\frac{1}{\xi_m^{2\mu-2I_{m+1}}} \left(
\frac{1}{AQ_2Q_1}+\cdots+
\frac{\xi_2\cdot\cdot\cdot\xi_{\Lambda-1}}{AQ_{\Lambda}Q_{\Lambda-1}}
\right)^{2\mu-j} \right]\right\}\Bigg|_{(all \xi_m\rightarrow1)}\times\\
\noalign{\bigskip}
 &&\times\left(\frac{1}{AQ_0Q_1}\right)^j
\left\{\binom{I_2}{\lambda}a^{2\lambda}_{2\lambda-j}\ Q^{4\lambda}_0
Q_1^{4(I_2-\lambda)}\right\} \label{maineq}
\end {eqnarray}

We introduced the abbreviation
 $$I_j = \mu - (i_j +i_{j+1} + \cdots +i_{\Lambda} ).$$
$\xi_m$ are independent variables and $D_{\xi}$ is the differential operator given as:

$$D_{\xi}=3/4\partial^2_{\xi}+3 \xi \partial^3_{\xi}+\xi^2 \partial^4_{\xi}$$

The asymptotic decomposition of the function $\mathcal{S}_{\Lambda}$
reads:

\begin {equation}
\mathcal{S}_{\Lambda}=\sum\limits_{\mu=0}^{\mathcal{J}}\;\frac{(-1)^{\mu}}{\mu!\;(2z^2\triangle^3)^{\mu}}\;
\triangle^{3\mu}\left\{\ms C(\Lambda)^{2\mu}\right\}_{2\mu, 2\mu}\ ,
\label{maineq2}
\end {equation}
since $\{\ms C(\Lambda)^{2\mu}\}_{2\mu, 2\mu} = (\Lambda)^{2\mu}_0.\
$ The quantity $z^2\triangle^3$ is finite in the continuum limit
$\triangle\rightarrow \infty$ and the factor $\triangle^{3\mu}$
ensures that only the leading term of $\left\{\ms
C(\Lambda)^{2\mu}\right\}_{2\mu, 2\mu}$ survives the continuum
limit. Due to the analytic form for $\left\{\ms
C(\Lambda)^{2\mu}\right\}_{2\mu, 2\mu}$ we can express
$\mathcal{S}_{\Lambda}$ in the continuum $\triangle\rightarrow 0$
limit as well as in asymptotic $\mu\rightarrow \infty$ limit. The
analytical evaluations of the lower $\mu$ terms and the relation for
the asymptotic terms is done in details in our article
\cite{paper2}. We evaluated the analytical result for the first
three terms (i.e. $\mu=1,2,3$). As an illustration we give the first
of them:

$$\left\{\ms C(\Lambda)^2\right\}_{2,2}=\frac{\mu!}{(2\mu)!}\sum_{k=2}^{\Lambda}\
Q_k^4\ b_k^2\ \textbf{J}(0,0;2,1;b_k/b_{k-1})$$
where
$$\textbf{J}(l,MIN;n,i_j;b_j/b_{j-1}) = \sum_{p=l}^{MIN}\ \binom{2i_j-l}{p-l}\ \binom{2i_j-1/2}{2i_j-p}\
\frac{n!(2i_j-p)!}{(n-2i_j-p)!}\ \left(\frac{b_j}{b_{j-1}}\right)^p
,$$ and $$b_j = \frac{1}{Q_{j+1}Q_j}+ \cdots +
\frac{1}{Q_{\Lambda}Q_{\Lambda}-1}.$$ The continuum limit
corresponds to the prescription:

$$k.\triangle\rightarrow x,\ \triangle.\; \Lambda \rightarrow \tau,$$

$$\sum_{k=2}^{\Lambda} \rightarrow \frac{1}{\triangle}\
\int_0^{\tau}\ d x,$$ When $\Lambda \rightarrow N,$ then $\tau
\rightarrow \beta,$ where $\beta$ is the the constant of the model,
and $N= \frac{\beta}{\triangle}.$

In the continuum limit we obtain:
$$Q_k \rightarrow 2\cosh(\gamma x)$$
$$b_k \rightarrow \frac{1}{\triangle \gamma}(\tanh(\gamma \tau)-\tanh(\gamma x))$$
where $\gamma = \sqrt{2b/c},$ $b$ and $c$ are parameters of the
model. The continuum limit of the relation (\ref{maineq2}) will be
called $S(a,b,c,\tau)$. We show the first nontrivial term $(\mu=1)$
of the three evaluated now:

\begin {equation}
\left\{\ms C^2(a,b,c,\tau)\right\}_{2,2} = \frac{3}{8
\gamma^3}\left[3\gamma \tau \tanh^2(\gamma \tau) +
 \tanh(\gamma \tau) - \gamma \tau\right]
\end {equation}

 For the calculated  higher terms we have  analytical formulas also
as results of algebraic evaluation by Mathematica\cite{wolf}. The
continuum function $S(a,b,c,\tau)$ for the first three nontrivial
contributions is shown in Fig. 1.

\begin{figure}
  \includegraphics[width=9cm]{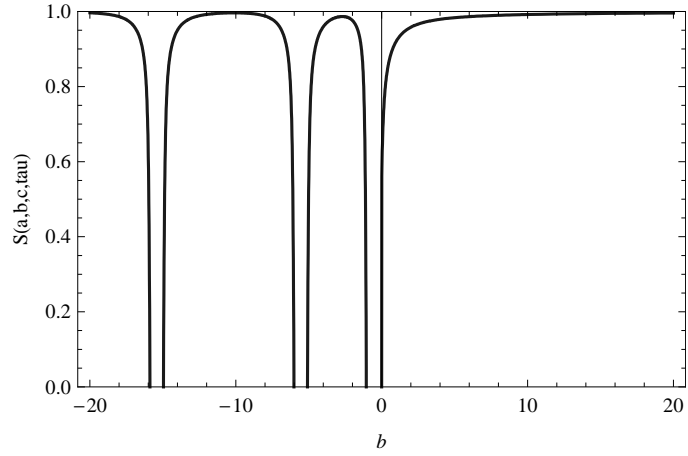}
  \caption {$b$ dependence of the continuum function $S(a,b,c,\tau)$ for fixed values $a=0.1, c=0.5, \tau=1.$ The first three nontrivial terms
 of the asymptotic series (\ref{maineq2}) were used.}
\end{figure}

The corresponding term for the Gelfand-Yaglom equation,
$-2\partial^2_{\mu}\ln(S(a,b,c,\tau))-4(\partial_{\mu}\ln(S(a,b,c,\tau)))^2,$
is shown in  Fig. 2.

\begin{figure}
  \includegraphics[width=9cm]{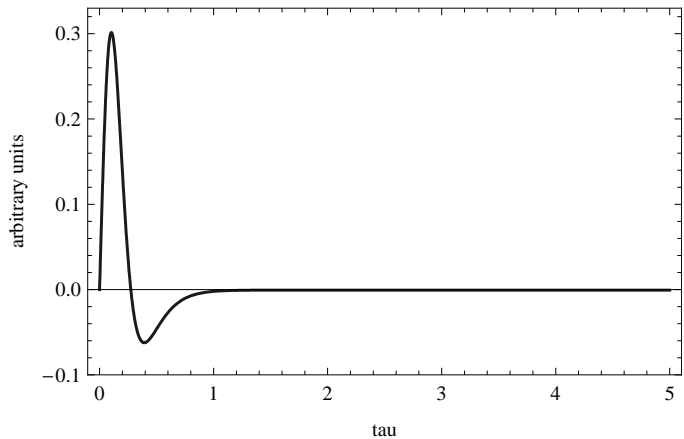}
 \caption{$\tau$ dependence of the continuum function
 $-2\partial^2_{\mu}\ln(S(a,b,c,\tau))-4(\partial_{\mu}\ln(S(a,b,c,\tau)))^2$ for fixed $a=0.1, b=5, c=0.5.$
 The first three nontrivial terms
 of the asymptotic series (\ref{maineq2}) were used.}
\end{figure}

In the limit $\mu\rightarrow \infty$, $\triangle$ fixed the terms
(\ref{maineq}) are divergent. For the leading divergent term for
$\mu \rightarrow \infty$ of the asymptotic series for
$\mathcal{S}_{\Lambda}$ (\ref{maineq2}) we find \cite{paper2}:

\begin {equation}
\frac{(-1)^{\mu}a^{\mu}}{\mu!\;c^{2\mu}}\; \triangle^{\mu}2^{\mu}\
\pochh{1/2}{2\mu}\left(\frac{\tanh{(\gamma
\tau)}}{\gamma}\right)^{2\mu}
\end {equation}

The series of this form is an asymptotic expansion of the parabolic
cylinder function of the index $-1/2$ and the argument
$$ z^{-1} =2 a \triangle\left(\frac{\tanh{(\gamma
\tau)}}{c \gamma}\right)^2\ .$$
For small $\mu$ the terms of the decomposition (\ref{maineq2}) don't fulfil
 such decomposition.

\section{conclusions}

We presented an analytical method of evaluation of the unconditional
 Wiener measure functional integral with a fourth order term in the action.
No simple analytical form of such an integral is known, perturbative
methods of evaluation are needed. Instead of a standard perturbative
procedure we expand the linear term of the action. Such integral is
an entire function of all remaining parameters with infinite radii
of convergence. We find an analytical result for the functional
integral in the form of the solution of the "generalized
Gelfand-Yaglom" (GGY) equation  for the case of an anharmonic
oscillator with positive coupling and no positivity requirement for
the quadratic term. We calculated the asymptotic solution of the GGY
equation up to third order in the coupling constant.

The same method of evaluation could be applied for the case of the
functional integral with the conditional Wiener measure for anharmonic
oscillator, with more interesting physical results. The approach
based on the GGY equation can be applied too. However in this case the
method is technically rather involved.
 An evaluation
for a simplified case when the endpoints are fixed to  zero is
presented in our article I \cite{paper1}. In this case the
functional integral is a propagator with coinciding endpoints well
known as Moeler's formula \cite{dem} in the case of  harmonic
oscillator. We evaluated the correction to  Moeler's formula
$S(\beta)$ up to the first nontrivial term and we found:
\begin {equation}
S(\beta)=1- \frac{3 a}{32c^2\gamma^3} \left\{-3\coth(\gamma\beta)+
2\gamma\beta\left[\coth^2(\gamma\beta)+\frac{1}{2\sinh^2(\gamma\beta)}\right]\right\}
\end {equation}
A more complete evaluation of the functional integral with
conditional Wiener measure and a more systematic study of the
anharmonic oscillator is in progress.

When the frequency $b(\tau)$ and the coupling constant $a(\tau)$ are
time dependent we obtain for the anharmonic oscillator by the time
slicing method described in the article an equation which in the
continuum limit reads:

\begin {equation}
\frac{\partial^2}{\partial \tau^2}F(\tau)+4\frac{\partial}{\partial
\tau}F(\tau)\, \frac{\partial}{\partial \tau}\ln{S(\tau)}=
F(\tau)\left[\frac{2b(\tau)}{c}-2\frac{\partial^2}{\partial
\tau^2}\ln{S(\tau)}-4\left(\frac{\partial}{\partial
\tau}\ln{S(\tau)}\right)^2\right], \label{rggye}
\end {equation}
where $S(\tau)$ can be evaluated by a method similar to that of Section $IV$.
The time dependence of the functions $a(\tau),\
b(\tau),$ instead of constants $a,\ b,$ does not complicate
summations over the index in individual time-slice intervals. The
nonlocal character of the result for $N-$ dimensional integral,
represented by the dependence of parabolic cylinder functions on two
summation indexes is represented by the function $\xi_i,$ introduced
in the evaluation procedure:
$$(1+b_{N-m-1}\triangle^2/c)(1+b_{N-m}\triangle^2/c)\xi^2_{N-m-1}=1$$
After the full recurrence procedure of the evaluation of the $N-$ dimensional integral, we obtain the result:
\begin {equation}
\mathcal{Z}_{N-1}^{cut}=
\left\{\prod\limits_{i=0}^{N-1}\frac{1}{\sqrt{2(1+b_{N-i}\triangle^2/c)\omega_i}}\right\}
S_{N-1},
\end {equation}
where $S_{N-1}$ is a function with structure unimportant for this moment.
The function $\omega_i$ is defined by the recurrence relation:
$$ \omega_i=1-\frac{\xi^2_{N-i}}{4\omega_{i-1}}.$$
We obtained  Eq. (\ref{rggye}) without detailed
evaluation of the function $S(\tau)$, by the same method as in Section $III$.
By substitution $$F(\tau)=\frac{y(\tau)}{S^2(\tau)}$$ we convert
 (\ref{rggye}) to selfadjoint \cite{kamke} equation:
\begin {equation}
y''(\tau)=\left(\frac{2b(\tau)}{c}\right)\ y(\tau).
\label{rg2}
\end {equation}

In the case of a linear harmonic oscillator with time dependent
frequency, we obtain for the function $F(\tau)$,  leading to the inverse square
root of the functional integral, the simple second order selfadjoint
equation discussed for the first time by Lewis \cite{lewis}. The time
evolution of  QM systems of a broader class of  time
dependent Hamiltonians was discussed by \v Samaj \cite{samaj}.
Equation (\ref{rggye}) belongs to the class of general linear second
order differential equations discussed exhaustively by Kamke
\cite{kamke}. Equation (\ref{rg2}) corresponds to its reduced normal
form with invariant $$I=-\frac{2b(\tau)}{c}.$$ In Kamke \cite{kamke}, the variable substitution and
function replacement leading to the differential equation of the second order with constant invariant are proven.
Therefore, as in the case of the harmonic oscillator, in the case of
anharmonic oscillator there exists a possibility to convert the
problem with time dependent frequency and coupling constant to the constant
coefficient linear equation of the second order problem.

Functional integral methods play an important role in quantum
mechanics. One has the tools to evaluate the mean values of
observables without a necessity to solve  equations of motion. For
example, the conditional measure functional integral in quantum
mechanics represents particle propagation. Its integrated
form over final positions describes unconditional functional
integral in quantum mechanics. In statistical physics, the same
quantity, after Wick rotation $it\rightarrow -\tau,$ represents the
partition function. In fact,  all such functional integrals offer
important information about physical quantities, e.g. spectrum of
Hamiltonian, mean values of observables, etc.

The anharmonic oscillator can be considered as $\phi^4$ theory in
$(1+0)$ dimensions and  can be regarded as a toy model for
understanding of QCD, as the anharmonic oscillator \cite{bender}
studied in the perturbative approach was. In our description we have a
possibility to study the case with positive mass, corresponding to the
anharmonic oscillator and, what is more interesting, the case with
negative frequency.
It will be interesting to analyze the analog of the GGY equation in
field theory also.

\vskip 1.3cm {\bf{Acknowledgements}}. This work was supported by
VEGA project No. 2/6074/26.

\appendix

\section{ Evaluation of the $N-$ dimensional integral}

A pedagogical evaluation of the integral is done in our article I
\cite{paper1}, here we recall the most important steps. We are going
to evaluate the $N-$ dimensional integral defined by the relation:
\begin {equation}
\mathcal{Z}_{N}=\int\limits _{-\infty}^{+\infty} \prod \limits
_{i=1}^N\left(\frac{d\varphi_i}{\sqrt{\frac{2\pi\triangle}{c}}}\right)
\exp\left\{-\sum\limits _{i=1}^N \triangle\left[c/2
\left(\frac{\varphi_i-\varphi_{i-1}}{\triangle}\right)^2
+b\varphi_i^2+a\varphi_i^4\right]\right\}\ . \label {pcf3}
\end {equation}

First, we rewrite the sum in exponential function in a form
convenient for consecutive integrations:
\begin {eqnarray}
\mathcal{L}_N& = &\triangle a \varphi_1^4\ + \
\frac{c}{\triangle}(1+\frac{b \triangle^2}{c}) \varphi_1^2\ - \
\frac{c}{\triangle} \varphi_1 \varphi_2 \nonumber \\ & + & \cdots \nonumber\\
& + & \triangle a \varphi_i^4\ + \ \frac{c}{\triangle}(1+\frac{b
\triangle^2}{c}) \varphi_i^2\ - \ \frac{c}{\triangle} \varphi_i
\varphi_{i+1} \\ & + & \cdots \nonumber\\ & + & \triangle a
\varphi_N^4\ + \ \frac{c}{\triangle}(1/2+\frac{b \triangle^2}{c})
\varphi_N^2\ . \nonumber
\end {eqnarray}
We expand the exponential factor containing terms linear in the
integration variable into Taylor's series. Using the integration
formula \cite{prud}:

\begin {equation}
\int_0^{\infty}\ x^{\alpha-1}\ \exp (-p x^2-q x)\ d x = \Gamma
(\alpha)(2 p)^{-\alpha/2} \exp\left(\frac{q^2}{8 p}\right)
D_{-\alpha}\left(\frac{q}{\sqrt{2p}}\right), \label {prudnik1}
\end {equation}
\noindent
we find for integration over the variable $\varphi_1$:
\begin {equation}
\mathcal{Z}_1  = \frac{1}{\sqrt{2\pi(1+b\triangle^2/c)}}\;
\sum\limits _{k_1=0}^{\infty}\; \frac{\left(\frac{\scriptstyle
c}{\scriptstyle
\triangle(1+b\triangle^2/c)}\right)^{k_1}}{(2k_1)!}\;(\varphi_2)^{2k_1}\;
\Gamma(k_1+1/2)\mathcal{D}_{-k_1-1/2}(z)\ , \label{z111}
\end {equation}
where we used the notation:
\begin {equation}
 \mathcal{D}_{-k_1-1/2}(z) = z^{k_1+1/2}
e^{\frac{\scriptstyle z^2}{\scriptstyle 4}}\; D_{-k_1-1/2}(z)\ ,\
z=\frac{c(1+b\triangle^2/c)}{\sqrt{2a\triangle^3}}\ .
\end {equation}
 The term $(\varphi_2)^{2k_1} $ in Eq.(\ref{z111}) will play active
r\^{o}le in the integration over the variable $\varphi_2$:
\begin {equation}
\mathcal{Z}_2^{loc} = \int\limits
_{-\infty}^{+\infty}\;\frac{d\varphi_2}{\sqrt{\frac{2\pi\triangle}{c}}}\;
(\varphi_2)^{2k_1}\;\exp\left\{-\triangle a \varphi_2^4\ - \
\frac{c}{\triangle}(1+\frac{b \triangle^2}{c}) \varphi_2^2 +
\frac{c}{\triangle} \varphi_2 \varphi_3 \right\}\ .
\end {equation}
\noindent Taking both integration steps together we have:
\begin {equation}
\mathcal{Z}_2 =
\left(\frac{1}{\sqrt{2\pi(1+b\triangle^2/c)}}\right)^2  \sum
\limits_{k_1,k_2=0}^\infty  \frac
     {\xi^{2k_1} \left(\frac{\scriptstyle
c}{\scriptstyle
\triangle(1+b\triangle^2/c)}\varphi_3^2\right)^{k_2}}
     {(2k_1)!(2k_2)!}\;
\Gamma(k_1+1/2)\mathcal{D}_{-k_1-1/2}(z)\;\Gamma(k_1+k_2+1/2)\mathcal{D}_{-k_1-k_2-1/2}(z)\
, \label{a7}
\end {equation}
\noindent where we used a new symbol:
$$\xi = \frac{1}{(1+b\triangle^2/c)}\ .$$ Eq. (\ref{a7}) is the result of both $\varphi_1$ and $\varphi_2$
integrations. We can repeat this procedure for other integration
variables $\varphi_3, \cdots, \varphi_{N-1}.$ For the integration
over variable $\varphi_N$ one has no linear term in the exponent,
therefore we don't expand anything and this last integration will
not add summation over the index $k_N$ to the final formula. We
have:
\begin {equation}
\mathcal{Z}_N^{loc} = \int\limits
_{-\infty}^{+\infty}\;\frac{d\varphi_N}{\sqrt{\frac{2\pi\triangle}{c}}}\;
(\varphi_N)^{2k_{N-1}}\;\exp\left\{-\triangle a \varphi_N^4\ - \
\frac{c}{\triangle}(1/2+\frac{b \triangle^2}{c}) \varphi_N^2
\right\}\ .
\end {equation}
and the result is:
\begin {equation}
\mathcal{Z}_N^{loc} =
\frac{\left(c(1/2+b\triangle^2/c)/\triangle\right)^{k_{N-1}}}{\sqrt{2\pi(1/2+b\triangle^2/c)}}\;
\Gamma(k_{N-1}+1/2)\mathcal{D}_{-k_{N-1}-1/2}(z_{N})\ . \nonumber
\end {equation}
Remember the difference in definitions of $z_N$ and $z_i,\ i=1, 2,
\cdots , N-1$:
$$z_{N} = \frac{c(1/2+b\triangle^2/c)}{\sqrt{2a\triangle^3}}$$ and
also the term
$$\left(c(1/2+b\triangle^2/c)/\triangle\right)^{k_{N-1}}\ ,$$ which
modify the definition of:
$$\xi_{N-1} = \sqrt{\frac{1}{(1+b\triangle^2/c)}}\ \sqrt{\frac{1}{(1/2+b\triangle^2/c)}}\ .
$$

For the $N-$ dimensional integral we obtain finally the exact
result:
\begin {equation}
\mathcal{Z}_{N} =
\left[2\pi(1+b\triangle^2/c)\right]^{-\frac{N-1}{2}}
\left[2\pi(1/2+b\triangle^2/c)\right]^{-1/2}
\sum\limits_{k_1,\cdots,k_{N-1}=0}^\infty \prod \limits _{i=1}^N \;
\left[ \frac{\left(\xi_i\right)^{2k_{i}}}{(2k_{i})!}
\Gamma(k_{i-1}+k_{i}+1/2)\mathcal{D}_{-k_{i-1}-k_{i}-1/2}\;(z_i)\right]\
,
\end {equation}
\noindent where $k_0 \equiv k_N \equiv 0,$ and $\xi_1 = \xi_2 =
\cdots = \xi_{N-2} = \xi,$ also $\xi_{N} =1\ ,$ and $z_1 = z_2 =
\cdots = z_{N-1} = z.$

\section{ Proof of the generalized Gelfand-Yaglom differential equation}

We rewrite the $N-$ dimensional integral given in Eqs. (\ref{ndim}),
(\ref{ndim1}) as follows:
$$\mathcal{Z}_N =
\frac{S_{N-1}(\triangle)}{\sqrt{\prod\limits^{N-1}_{i=0}2(1+b\triangle^2/c)\omega_i}},$$
where $$\omega_i=1-\frac{A^2}{\omega_{i-1}}, \
\omega_0=\frac{1/2+b\triangle^2/c}{1+b\triangle^2/c}, \
A=\frac{1}{2(1+b\triangle^2/c)}.$$
The value of the functional integral in the
continuum limit is formally defined by $$\mathcal{Z} = \lim_{N
\rightarrow \infty}\mathcal{Z}_N$$

Let us define the function
\begin {equation}
F_k =
\frac{\prod\limits^{k}_{i=0}2(1+b\triangle^2/c)\omega_i}{S^2_{k}}
\label{gjf1}
\end {equation}
Here $S_{k}$ is the function $S_{N}(\triangle)$, with $N$ replaced
by $k$ and $\triangle$ is fixed. Let us stress the relation between
the $N-$ dimensional integral and the function $F_N$:
$$\mathcal{Z}_N = \frac{1}{\sqrt{F_N}}$$
The aim of the Gelfand-Yaglom construction is to find in the
continuum limit such a differential equation in variable $\tau \sim
k.\triangle$  that its solution is connected to the continuum
functional integral by relation:
$$\mathcal{Z}(\beta) =\frac{1}{\sqrt{F(\beta)}},\ \beta \sim N.\triangle.$$

In the spirit of the Gelfand-Yaglom construction we are going to
express the relation for $F_{k+1}$ by help of $F_{k}$ and $F_{k-1}$.
We have:

\begin {eqnarray}
F_{k+1}&=&\frac{\prod\limits^{k+1}_{i=0}2(1+b\triangle^2/c)\omega_i}{S^2_{k+1}}=
\frac{2(1+b\triangle^2/c)\omega_{k+1}\prod\limits^{k}_{i=0}2(1+b\triangle^2/c)\omega_i}{S^2_{k+1}}=
\\
\noalign{\vskip8pt}
&=&\frac{2(1+b\triangle^2/c)\prod\limits^{k}_{i=0}2(1+b\triangle^2/c)\omega_i}{S^2_{k+1}}-
\frac{\prod\limits^{k-1}_{i=0}2(1+b\triangle^2/c)\omega_i}{S^2_{k+1}}\nonumber
\label{gyeq3}
\end {eqnarray}

Regarding to the definition of the function $F_k$, in Eq.
(\ref{gjf1}), we have:
\begin {equation}
F_{k+1}=2(1+b\triangle^2/c)F_k\ \frac{S^2_{k}}{S^2_{k+1}}-
F_{k-1}\frac{S^2_{k-1}}{S^2_{k+1}} \label{gyeq4}
\end {equation}
After some algebra we find:
\begin {eqnarray}
& &F_{k+1}-2F_k+F_{k-1} -
2(F_k-F_{k-1})\left(\frac{S_k^2-S_{k+1}^2}{S_{k+1}^2}\right)=
\frac{2b\triangle^2}{c} F_k\
-F_{k-1}\left(\frac{S^2_{k-1}-2S_k^2+S^2_{k+1}}{S^2_{k+1}}\right)\nonumber \\
\noalign{\vskip8pt} &+& \frac{2b\triangle^2}{c}
F_k\left(\frac{S^2_{k}-S^2_{k+1}}{S^2_{k+1}}\right)
\label{gyeq6}
\end {eqnarray}
We need not to know the structure of functions $S_k$ to derive
the identities:
\begin {equation}
S^2_k-S^2_{k+1}=-2S_{k+1}(S_{k+1}-S_k) + (S_{k+1}-S_k)^2
\end {equation}
and
\begin {eqnarray}
S^2_{k-1}-2S_k^2+S^2_{k+1}&=&2(S_{k+1}-S_k)^2 +2S_{k+1}(S_{k+1}-2S_k+S_{k-1})-\\
\noalign{\vskip8pt}
&-&4(S_{k+1}-S_k)(S_{k+1}-2S_k+S_{k-1})+(S_{k+1}-2S_k+S_{k-1})^2\
.\nonumber
\end {eqnarray}
Inserting these identities into Eq. (\ref{gyeq6}) we find a
difference equation which, divided by $\triangle^2,$ takes the form:
\begin {eqnarray}
&&\frac{F_{k+1}-2F_{k}+F_{k-1}}{\triangle^2}+
4\frac{F_{k}-F_{k-1}}{\triangle}\frac{S_{k+1}-S_{k}}{\triangle\;
S_{k}} = F_k\left[2b/c -2\frac{S_{k+1}-2S_{k}+S_{k-1}}{\triangle^2\;
S_{k}}-2\left(\frac{S_{k+1}-S_{k}}{\triangle\;
S_{k}}\right)^2\right] \\
\noalign{\vskip8pt}
 &+&\triangle \mathcal{O}_1+\triangle^2 \mathcal{O}_2\ ,\nonumber
\end {eqnarray}
\noindent where
\begin {eqnarray}
\mathcal{O}_1 &=&- \frac{4b}{c}F_k\left(\frac{S_{k+1}-S_k}{\triangle
S_{k+1}}\right)
+2\frac{(F_k-F_{k-1})}{\triangle}\left[\frac{b}{c}+\left(\frac{S_{k+1}-S_k}{\triangle
S_{k+1}}\right)^2\right] +4F_{k-1}\left(\frac{S_{k+1}-S_k}{\triangle
S{k+1}}\right)\left(\frac{S_{k+1}-2S_k+S_{k-1}}{\triangle^2
S_{k+1}}\right) \nonumber
\\ \noalign{\vskip8pt}
\mathcal{O}_2 &=& \frac{2b}{c}F_k\left(\frac{S_{k+1}-S_k}{\triangle
S_{k+1}}\right)^2-
F_{k-1}\left(\frac{S_{k+1}-2S_k+S_{k-1}}{\triangle^2
S_{k+1}}\right)^2
\end {eqnarray}
 In the continuum limit $\triangle \rightarrow 0$ we replace $k . \triangle$ by $\tau.$ Under the condition
\begin {equation}
\lim_{\triangle \rightarrow 0}(\triangle \ \mathcal{O}_1+\triangle^2
\ \mathcal{O}_2) = 0\ ,
\end {equation}
 we  obtain the differential equation:
\begin {equation}
\frac{\partial^2}{\partial \tau^2}F(\tau)+4\frac{\partial}{\partial
\tau}F(\tau)\, \frac{\partial}{\partial \tau}\ln{S(\tau)}=
F(\tau)\left[\frac{2b}{c}-2\frac{\partial^2}{\partial
\tau^2}\ln{S(\tau)}-4\left(\frac{\partial}{\partial
\tau}\ln{S(\tau)}\right)^2\right] \label{ggye}
\end {equation}
The initial conditions for $S(\tau)$ continuous and finite in $\tau
= 0$ are:
\begin {eqnarray}
F(0) &=& \frac{1}{S^2(0)},
\label{gyeqa2} \\
\frac{\partial}{\partial \tau}F(0) &=& \lim_{\triangle\rightarrow
0}\
\frac{F_1-F_0}{\triangle}=-\left(\frac{1}{S^2(0)}\right)'.\nonumber
\end{eqnarray}

\section{Evaluation of the leading part of $S_N$}

In this appendix we  evaluate the finite range $k_i$ summations by the
recurrence method.
 Let us start with summation over the index
$k_{N-1}$ of the Eq.(\ref{fin1}). The finite sum to be done is:
\begin {eqnarray}
\mathcal{Z}_1^{cut} &=& \sum \limits_{k_{N-1}=0}^{K_0} \;
\left[\frac{1}{\sqrt{2\pi(1+b\triangle^2/c)}} \;
\frac{\left(\xi^2/\omega_0\right)^{k_{N-1}}}{(2k_{N-1})!}
\Gamma(k_{N-2}+k_{N-1}+1/2)\mathcal{D}_{-k_{N-2}-k_{N-1}-1/2}\;(z)\right]\
 \nonumber \\ \noalign{\vskip8pt}
& & \left[\frac{1}{\sqrt{2\pi(1+b\triangle^2/c)\omega_0}} \;
\Gamma(k_{N-1}+1/2)\mathcal{D}_{-k_{N-1}-1/2}\;(z_0)\right]
\label{app2}
\end {eqnarray}
Let us recall the dependence of the function $\mathcal{D}$ on the
parabolic cylinder function $D$:
$$\mathcal{D}_{-m-1/2}(z) = z^{m+1/2}\;
e^{\frac{\scriptstyle z^2}{\scriptstyle 4}}\; D_{-m-1/2}(z)$$ and
also the definitions of the variables $\xi$, $\omega_0$, $z$ and
$z_0$:
\begin {eqnarray*}
z & = & \frac{c(1+b\triangle^2/c)}{\sqrt{2a\triangle^3}},\
z_0 = \frac{c(1/2+b\triangle^2/c)}{\sqrt{2a\triangle^3}}, \\ \noalign{\vskip8pt}
\xi & =&  \frac{1}{1+b\triangle^2/c}, \
\omega_0 = \frac{1/2+b\triangle^2/c}{1+b\triangle^2/c}\ .
\end {eqnarray*}
 We use the asymptotic Poincar\' e-type
expansion of the parabolic cylinder function, which for
$\mathcal{D}$ means:
\begin {equation}
\mathcal{D}_{-k_n-1/2}(z_0)\; \equiv \;
z_0^{k_n+1/2}\;e^{z_0^2/4}\;D_{-k_n-1/2}(z_0)\;=
\;\sum\limits_{j=0}^{\mathcal{J}}\; (-1)^j
\;\frac{\pochh{k_n+1/2}{2j}}{j!\;(2z_0^2)^j}
+\varepsilon_{\mathcal{J}}(k_n, z_0)
\end {equation}
In the last relation, $\mathcal{J}$ denotes the number of  terms of
the asymptotic expansions convenient to take into account. We apply
asymptotic expansion for the function
$\mathcal{D}_{-k_{N-1}-1/2}\;(z_0)$ in Eq.(\ref{app2}). In the
truncated sum we interchange the order of the finite summations over
indices $k_{N-1}$ and $j$. We replace $\mathcal{D} $ by $D$,
therefore a corresponding power of the variable $z$ will play an
important role. In this way we obtain the relation:
\begin {eqnarray}
\mathcal{Z}_1^{cut}
&=&\frac{\Gamma{(1/2)}}{\sqrt{2\pi(1+b\triangle^2/c)\omega_0}}\;\sum\limits_{j=0}^{\mathcal{J}}\;
\frac{(-1)^j}{j!\;(2z_0^2)^j}\;
\frac{\exp{(z^2/4)}}{\sqrt{2\pi(1+b\triangle/c)}} z^{k_{N-2}+1/2}
\\&& \sum\limits_{k_{N-1}=0}^{K_0}\frac{(\frac{\scriptstyle z\
\xi^2}{\scriptstyle 4\omega_0})^{k_{N-1}}}{(k_{N-1})!}\;
\Gamma(k_{N-1}+k_{N-2}+1/2)\;\pochh{k_{N-1}+1/2}{2j}\;
D_{-k_{N-1}-k_{N-2}-1/2}\;(z), \nonumber \label{z1}
\end {eqnarray}
where we simplified the calculations by identities:
$$(2k)! = 2^{2k}\; k!\; \pochh{1/2}{k},\ \Gamma{(k+1/2)}=\Gamma{(1/2)}\pochh{1/2}{k}.$$

Let us study in detail the sum
\begin {equation}
\sum\limits_{k_{N-1}=0}^{K_0}\frac{(\frac{z\
\xi^2}{4\omega_0})^{k_{N-1}}}{(k_{N-1})!}\;
\Gamma(k_{N-1}+k_{N-2}+1/2)\;\pochh{k_{N-1}+1/2}{2j}\;
D_{-k_{N-1}-k_{N-2}-1/2}\;(z)
\end {equation}
This sum is uniformly convergent, therefore we can extend the
summation up to infinity by adding the corresponding terms, which
appear also in the remainder with opposite sign.
 To be able to provide the sum over the index
$k_{N-1}$, we must modify the Pochhammer symbol
$$\pochh{k_{N-1}+1/2}{2j}=(k_{N-1}+1/2).\cdots .(k_{N-1}+1/2+2j-1)\
.$$ We see, that this object is a polynomial in the variable
$k_{N-1}$ of the $2j-th$ order. We rewrite the polynomial in another
form:
$$\pochh{k_{N-1}+1/2}{2j}\; = \;
\sum\limits_{i=0}^{\min{(2j,k_{N-1})}}a_i^{2j}\frac{(k_{N-1})!}{(k_{N-1}-i)!}$$
The coefficients $a_i^{2j}$ are given by
recurrence procedure from the relation:
\begin {equation}
\sum\limits_{k_{N-1}=0}^{K_0}\frac{(k_{N-1}+1/2)_{2j}}{(k_{N-1})!}\;f(k_{N-1})\;
=\sum\limits_{i=0}^{2j}\;a_i^{2j}\;
\sum\limits_{k_{N-1}=i}^{K_0}\frac{1}{(k_{N-1}-i)!}\;f(k_{N-1})
\end {equation}
From the above definition, we find the recurrence equation:
\begin {equation}
a^k_i=(k-1/2+i)a^{k-1}_i\; + a^{k-1}_{i-1}
\end {equation}
while the initial conditions are: $$a_j^j=1\;,\ a^j_0 =
\pochh{1/2}{j}\; ,\ a^j_{j+1}=0$$ The solution of this recurrence
equation is:
\begin {equation}
a_i^{j}\; = \;\binom{j}{i}\frac{\pochh{1/2}{j}}{\pochh{1/2}{i}}
\end {equation}
Inserting all these replacements into Eq.(\ref{z1}), with help of
the identity
$$\Gamma(k_{N-2}+k_{N-1}+1/2) = \Gamma(k_{N-2}+i+1/2)\;
\pochh{k_{N-2}+i+1/2}{k_{N-1}-i}$$ after some algebra, introducing a
new summation index $k=k_{N-1} - i,$ we obtain the formula:
\begin {eqnarray}
\mathcal{Z}_1^{cut}
&=&\frac{z^{k_{N-2}+1/2}}{\sqrt{2(1+b\triangle^2/c)\omega_0}}\;\sum\limits_{j=0}^{\mathcal{J}}\;
\frac{(-1)^j}{j!\;(2z^2_0)^j}\;
\frac{e^{z^2/4}}{\sqrt{2\pi(1+b\triangle^2/c)}}
\sum\limits_{i=0}^{2j}\;a_i^{2j}\\
&\times& \Gamma(k_{N-2}+i+1/2)\;\left(\frac{z\
\xi^2}{4\omega_0}\right)^i\;\sum\limits_{k=0}^{K_0} \frac{(\frac{z\
\xi^2}{4\omega_0})^k}{k!}\;(k_{N-2}+i+1/2)_{k}\;
D_{-k_{N-2}-i-1/2-k}\;(z)\nonumber \quad
\end {eqnarray}
In the above relation we extend  summation over the index $k$ up to
infinity. The sum over $k$ is now prepared for application of the
identity:
\begin {equation}
e^{x^2/4}\sum\limits_{k=0}^{\infty}\;
\frac{\pochh{\nu}k}{k!}\;t^k\;D_{-\nu-k}(x)\;=\;e^{(x-t)^2/4}\;D_{-\nu}\;(x-t)
\end {equation}
The result of the first  recurrence step, replacing $D$ by
$\mathcal{D}$, reads:
\begin {equation}
\mathcal{Z}_1^{cut} =\frac{1}{\sqrt{2(1+b\triangle^2/c)\omega_0}}
\frac{(\omega_1)^{-k_{N-2}}}{\sqrt{2\pi(1+b\triangle^2/c)\omega_1}}
\sum\limits_{j=0}^{\mathcal{J}}\frac{(-1)^j}{j!\;(2z_0^2)^j}
\sum\limits_{i=0}^{2j}a_i^{2j}\left(\frac{\xi^2}{4\omega_0 \omega_1}\right)^i
\Gamma(k_{N-2}+i+1/2)\mathcal{D}_{-k_{N-2}-i-1/2}\;(z_1)
\label{rez11}
\end {equation}
where $$z_1=z\left(1-\xi^2/(4 \omega_0)\right),\; \omega_1 = \frac{z_1}{z} = 1-\xi^2/(4 \omega_0)\; .$$


For the following summation over the index $k_{N-2} $ we have:
\begin {eqnarray}
\mathcal{Z}_2 &=& \frac{1}{\sqrt{2(1+b\triangle^2/c)\omega_0}}
\frac{1}{\sqrt{2(1+b\triangle^2/c)\omega_1}} \\
& \times & \sum\limits_{k_{N-2}=0}^\infty \;
\left[\frac{1}{\sqrt{2\pi(1+b\triangle^2/c)}} \;
\frac{\left(\frac{\scriptstyle \xi^2}{\scriptstyle
\omega_1}\right)^{k_{N-2}}}{(2k_{N-2})!}
\Gamma(k_{N-3}+k_{N-2}+1/2)\mathcal{D}_{-k_{N-3}-k_{N-2}-1/2}\;(z)\right]\nonumber\\
& \times &
\sum\limits_{j=0}^{\mathcal{J}}\;\frac{(-1)^j}{j!\;(2z^2)^j}\;
\sum\limits_{i=0}^{2j}\;\left(1\right)_i^{2j}\; \left(\frac{\xi^2}{4
\omega_0 \omega_1}\right)^i\;
\Gamma(k_{N-2}+i+1/2)\mathcal{D}_{-k_{N-2}-i-1/2}\;(z_1)\ ,\nonumber
\label{app4}
\end {eqnarray}
where
\begin {equation}
\left(1\right)_i^{2j} = \frac{1}{\omega_0^{2j}}a_i^{2j}
\label{c10star}
\end {equation}
will define the first step of the new recurrence relation.

Now, due to the uniform convergence of the sum over the index
$k_{N-2}$ we will evaluate the leading part of $\mathcal{Z}_2$ as a
finite sum over index $k_{N-2}$. In the finite sum, we use the
asymptotic Poincar\' e-type expansion of the parabolic cylinder
function $\mathcal{D}_{-k_{N-2}-i-1/2}\;(z_1)$. We have then in the
relation for $\mathcal{Z}_2^{cut}$  finite summations only and we
change the order of the sums. We have:

\begin {eqnarray}
\mathcal{Z}_2^{cut}&=&
\frac{1}{\sqrt{2(1+b\triangle^2/c)\omega_0}}\;
\frac{1}{\sqrt{2(1+b\triangle^2/c)\omega_1}}
\frac{1}{\sqrt{2\pi(1+b\triangle^2/c)}} \\ \noalign{\vskip8pt}
& \times &
\sum\limits_{j=0}^{\mathcal{J}}\;\frac{(-1)^j}{j!\;(2z^2)^j}\;
\sum\limits_{i=0}^{2j}\;\left(1\right)_i^{2j}\;
\left(\frac{A^2}{\omega_0\omega_1}\right)^i\;
\sum\limits_{l=0}^{\mathcal{J}}\;\frac{(-1)^l}{l!\;(2z^2)^l}
\frac{1}{\omega_1^{2l}}\; e^{z^2/4}\ z^{k_{N-3}+1/2}\nonumber\\ \noalign{\vskip8pt}
 & \times & \sum\limits_{k_{N-2}=0}^{N_0} \;
\frac{\left(\frac{z\ \xi^2}{4 \omega_1}\right)^{k_{N-2}}}{k_{N-2}!}
\Gamma(k_{N-3}+k_{N-2}+1/2)\ D_{-k_{N-3}-k_{N-2}-1/2}\;(z)\
\pochh{k_{N-2+1/2}}{2l+i}\; ,\nonumber
\end {eqnarray}
where $A = \xi/2$. Summing over the index $k_{N-2}$ as in the first
recurrence step, we have:

\begin {eqnarray}
\mathcal{Z}_2^{cut}&=&
\frac{1}{\sqrt{2(1+b\triangle^2/c)\omega_0}}\;
\frac{1}{\sqrt{2(1+b\triangle^2/c)\omega_1}}
\frac{(\omega_2)^{-k_{N-3}}}{\sqrt{2(1+b\triangle^2/c)\omega_2}}
\sum\limits_{j=0}^{\mathcal{J}}\;\sum\limits_{l=0}^{\mathcal{J}}\;
\frac{(-1)^{j+l}}{j!l!(2z^2)^{j+l}}\;\frac{1}{\omega_1^{2l}}\label{acc}\\ \noalign{\vskip8pt}
& \times & \sum\limits_{i=0}^{2j}\;\left(1\right)_i^{2j}\;
\left(\frac{A^2}{\omega_0\omega_1}\right)^i\;
\sum\limits_{p=0}^{2l+i}\;a_p^{2l+i}\left(\frac{A^2}{\omega_1\omega_2}\right)^p\;
\Gamma(k_{N-3}+p+1/2)\mathcal{D}_{-k_{N-3}-p-1/2}(z_2)\; ,\nonumber
\end {eqnarray}
where one defines  new variables: $$z_2 = z \omega_2,\; \omega_2 =
1-\frac{A^2}{\omega_1}\; .$$

The summations over indices $j,l$ is done as follows:

\begin {eqnarray}
&&\sum\limits_{j=0}^{\mathcal{J}}\;\sum\limits_{l=0}^{\mathcal{J}}\;
\frac{(-1)^{j+l}}{j!l!(2z^2)^{j+l}}f(l+j)g(j)h(l) =\label{remrem}\\ \noalign{\vskip8pt}
&=&\sum\limits_{\mu=0}^{\mathcal{J}}\frac{(-1)^{\mu}f(\mu)}{\mu!(2z^2)^{\mu}}\;
\sum\limits_{j=0}^{\mu}\binom{\mu}{j}g(j)h(\mu-j)+
\sum\limits_{\mu=\mathcal{J}+1}^{2\mathcal{J}}\frac{(-1)^{\mu}f(\mu)}{(2z^2)^{\mu}}\;
\sum\limits_{j=\mu-\mathcal{J}}^{\mathcal{J}}\frac{g(j)}{j!}\frac{h(\mu-j)}{(\mu-j)!}\nonumber
\end {eqnarray}
The first term on the right hand side of the above relation will
contribute to the leading part of Eq.(\ref{acc}), while the second
term, where index $\mu >\mathcal{J}$, will contribute to the
remainder, due to the term $z^{-2\mu}\sim N^{-3\mu}$ which may be
made as small as possible by choosing  $\mathcal{J}$ properly.
Interchanging the order of summations:

$$\sum\limits_{i=0}^{2j}\;\sum\limits_{p=0}^{2\mu-2j+i}\;
\rightarrow \sum\limits_{p=0}^{\mu}\; \sum\limits^{2j}_{i=\max{[0,\;
p-2\mu+2j]}}$$ we find the result of the second recurrence step:

\begin {eqnarray}
\mathcal{Z}_2^{cut}&=&
\frac{1}{\sqrt{2(1+b\triangle^2/c)\omega_0}}\;
\frac{1}{\sqrt{2(1+b\triangle^2/c)\omega_1}}
\frac{(z/z_2)^{k_{N-3}}}{\sqrt{2(1+b\triangle^2/c)\omega_2}}\\ \noalign{\vskip8pt}
&\times&
\sum\limits_{\mu=0}^{\mathcal{J}+\mathcal{L}}\frac{(-1)^{\mu}}{\mu!(2z^2)^{\mu}}\;
\sum\limits_{p=0}^{2\mu}\;\left(2\right)_p^{2\mu}\left(\frac{A^2}{\omega_1\omega_2}\right)^p
\Gamma(k_{N-3}+p+1/2)\mathcal{D}_{-k_{N-3}-p-1/2}(z_2),\nonumber
\end {eqnarray}
where the second recurrence step of the function $(2)_p^{2\mu}$ is
defined by:
\begin {equation}
\left(2\right)_p^{2\mu}=
\sum\limits_{j=0}^{\mu}\;\binom{\mu}{j}\frac{1}{\omega_1^{2\mu-2j}}
\sum\limits^{2j}_{i=\max{[0,\; p-2\mu+2j]}}
\left(\frac{A^2}{\omega_0\omega_1}\right)^i\;
\left(1\right)_i^{2j}\; a_p^{2\mu-2j+i}
\end {equation}

We can see that after  $\Lambda$ recurrence steps the result of the
$\Lambda$ summations over the indices $k_i$ can be
read\cite{paper1}:
\begin {eqnarray}
\mathcal{Z}_{\Lambda}^{cut}&=&
\left\{\prod\limits_{i=0}^{\Lambda}\frac{1}{\sqrt{2(1+b\triangle^2/c)\omega_i}}\right\}
(\omega_{\Lambda})^{-k_{N-1-\Lambda}} \label {app11}\\ \noalign{\vskip8pt}
 &\times&
\sum\limits_{\mu=0}^{\mathcal{J}}\frac{(-1)^{\mu}}{\mu!(2z^2)^{\mu}}\;
\sum\limits_{p=0}^{2\mu}\;\left(\Lambda\right)_p^{2\mu}
\left(\frac{A^2}{\omega_{\Lambda-1}\omega_{\Lambda}}\right)^p
\Gamma(k_{N-\Lambda-1}+p+1/2)\mathcal{D}_{-k_{N-\Lambda-1}-p-1/2}(z_{\Lambda})\; .\nonumber
\end {eqnarray}
We have evaluated the recurrence relations: $$z_{\Lambda} = z
\omega_{\Lambda}, \; \omega_{\Lambda} =
1-\frac{A^2}{\omega_{\Lambda-1}},\; \omega_0 = \frac{1/2 +
b\triangle^2/c}{1 + b\triangle^2/c}\; , A=\frac{\xi}{2}\ ,$$ and
introduced the recurrence definition for the function
$\left(\Lambda\right)_p^{2\mu}$:
\begin {equation}
\left(\Lambda\right)_p^{2\mu}=
\sum\limits_{j=0}^{\mu}\;\binom{\mu}{j}\frac{1}{\omega_{\Lambda-1}^{2\mu-2j}}
\sum\limits^{2j}_{i=\max{[0,\; p-2\mu+2j]}}
\left(\frac{A^2}{\omega_{\Lambda-2}\omega_{\Lambda-1}}\right)^i\;
\left(\Lambda-1\right)_i^{2j}\; a_p^{2\mu-2j+i}
\label{recla}
\end {equation}
where the recurrence procedure begins from $\left(1\right)_i^{2j}$
given in Eq. (\ref{c10star}).

 After the last
recurrence step, for $\Lambda=N-1$, we are left with the relation of
the form (\ref {app11}) where $k_{N-\Lambda-1}\equiv 0$ and the
index of the $\mathcal{D}$ function is only $-p-1/2$. We expand the
$D_{-p-1/2}(z_{N-1})$ as in all previous recurrence steps and  find
the relation:
\begin {eqnarray}
\mathcal{Z}_{N-1}^{cut}&=&
\left\{\prod\limits_{i=0}^{N-1}\frac{1}{\sqrt{2(1+b\triangle^2/c)\omega_i}}\right\}
\sum\limits_{\mu=0}^{\mathcal{J}}\frac{(-1)^{\mu}}{\mu!(2z^2)^{\mu}}\nonumber \\
 &\times&
\sum\limits_{l=0}^{\mu}\;\binom{\mu}{l}\frac{1}{\omega_{N-1}^{2l}}
\sum\limits^{2l}_{i=0}
\left(\frac{A^2}{\omega_{N-2}\omega_{N-1}}\right)^i\;
\left(N-1\right)_i^{2j}\;\pochh{1/2}{2l+i}
\end {eqnarray}
Following the definition of the $a_i^j$ symbols, we have:
$$\pochh{1/2}{2l+i} =  a_0^{2l+i}$$ Then in the last part of the
preceding equation we read:
\begin {equation}
\sum\limits_{l=0}^{\mu}\;\binom{\mu}{l}\frac{1}{\omega_{N-1}^{2l}}
\sum\limits^{2l}_{i=0}
\left(\frac{A^2}{\omega_{N-2}\omega_{N-1}}\right)^i\;
\left(N-1\right)_i^{2j}\;a_0^{2l+i}\; = \; \left(N\right)_0^{2\mu}
\end {equation}

Following the calculations done in this Appendix, we conclude, that
it is possible to realize  summations in the exact formula for the
$N-$ dimensional integral at least by help of  asymptotic expansions
of the parabolic cylinder functions. It is possible to provide the
continuum limit of  our result and there are no additional terms
contributing to the result in the continuum limit. The result reads
\begin {equation}
\mathcal{Z}_{N-1}^{cut}=
\left\{\prod\limits_{i=0}^{N-1}\frac{1}{\sqrt{2(1+b\triangle^2/c)\omega_i}}\right\}
\sum\limits_{\mu=0}^{\mathcal{J}}\frac{(-1)^{\mu}}{\mu!(2z^2)^{\mu}}\left(N\right)_0^{2\mu}
\label{vysl1}
\end {equation}
This expression is sufficient for the  calculation of the continuum
 unconditional Wiener measure functional integral by the Gelfand-Yaglom
procedure leading to the differential equation of the second order.
The second part of the relation (\ref{vysl1}) represents the
expansion of an unknown function.

In the following calculation  the key role play  objects $\omega_i$
defined by the recurrence relation $$\omega_i\; =\;
1-\frac{A^2}{\omega_{i-1}}$$ with the first term
$$\omega_0 \; =\; 1/2 + B,$$ where $$B= \frac{b\triangle^2/c}{2(1+b\triangle^2/c)}.$$
$\omega_i$  defined in this way are represented by  continued
fractions. The continued fraction can be represented by a simpler
relation as the solution of its $n-th$ convergent problem of
continued fractions \cite{findif}. Let us shortly explain this
procedure.

Let us have a continued fraction of the form:
$$ \omega = a_1+\frac{\scriptstyle b_1}{\scriptstyle a_2+
\frac{\scriptstyle b_2}{\scriptstyle a_3+\cdots}}$$ The $n-th$
convergent is defined as
$$\omega_n = \frac{p_n}{q_n}$$
where $p_n$ and $q_n$ are given by equations:
\begin {eqnarray}
p_n &=& a_n \;p_{n-1}+b_n \; p_{n-2}\nonumber\\
q_n &=& a_n \;q_{n-1}+b_n \; q_{n-2}\nonumber\\
\end {eqnarray}
Solutions of these recurrence equations have the form:
\begin {eqnarray}
p_n &=& \tilde{w}_1 \rho_1^n + \tilde{w}_2 \rho_2^n\nonumber\\
q_n &=& w_1 \rho_1^n + w_2 \rho_2^n \nonumber\\
\end {eqnarray}
where $\rho_{1,2}$ are  solutions of the characteristic equation, in
our case a homogenous one:
$$ \rho^2 -a_n\rho-b_n = 0 $$
For the continued fraction in question we have:
$$a_n=1,\  b_n=-A^2\ ,$$ and the solution of the characteristic equation
is:
 $$\rho_{1,2} = \frac{1}{2}(1\pm \sqrt{1-4A^2}\;)$$
 The constants $\tilde{w}_1$ and $w_1$ are fixed by $\omega_0$ and
 $\omega_1$ terms, that adjust the initial conditions:
\begin {eqnarray*}
p_0 &=& 1+2B\\ p_1 &=& 1+2B-A^2\\ q_0 &=& 2\\ q_1 &=& 1+2B
\end {eqnarray*}
The $n-th$ convergent method solution is completed by the
relations:
\begin {eqnarray*}
\tilde{w}_{1,2} &=& \frac{1}{2}\left[(1+2B)\pm
\left(\sqrt{1-4A^2}+\frac{2B}{\sqrt{1-4A^2}}\right)\right]\ ,\\
w_{1,2} &=& 1 \pm \frac{2B}{\sqrt{1-4A^2}}\ ,
\end {eqnarray*}
the particular characteristic follows from the above solution:
$$p_n\; = \; q_{n+1}$$
which simplifies our calculation significantly. In forthcoming
calculations we have introduced  more convenient variables:
$$ Q_{i}=\frac{q_i}{A^i} = w_1\left(\frac{\rho_1}{A}\right)^i +
w_2\left(\frac{\rho_2}{A}\right)^i$$ and also  performed the
replacement:
$$ \frac{A^2}{\omega_{k-1}^2} = \frac{Q_{k-1}^2}{Q_k^2}\ . $$

In what follows, we:

\noindent - replace the summation index $i$ by the summation index
$j$ defined by $i=2\lambda - j\ .$

\noindent - interchange the order of summations over indexes $j$ and
$\lambda$.

\noindent - rewrite the recurrence relation Eq. (\ref{recla}) as
follows:
\begin {eqnarray}
&&(A Q_{\Lambda} Q_{\Lambda-1})^{2\mu-p}
\left(\Lambda\right)_{2\mu-p}^{2\mu} = \nonumber
\\ \noalign{\bigskip}
&=&\sum\limits_{j=0}^p\;\frac{a^{2\mu-j}_{2\mu-p}}{(A Q_{\Lambda}
Q_{\Lambda-1})^{p-j}}\sum\limits^{\mu}_{\lambda=[\frac{j+1}{2}]}
\left(A Q_{\Lambda-2}Q_{\Lambda-1}\right)^{2\lambda-j}\;
\left(\Lambda-1\right)_{2\lambda-j}^{2\lambda}\;
\binom{\mu}{\lambda}(Q_{\Lambda-1}^4)^{\mu-\lambda}\ ,\ p\in <0,
2\mu>\ ,
\end {eqnarray}
the recurrence starts from the term (\ref{c10star})

\section{Evaluation of the remainder to the leading part of $S_N$}

Let us describe the evaluation of the remainder to the leading part
of the sum calculated in  Appendix C. The remainder consists of the
infinite sum of the original series, the finite sum over the
remainder of the Poincar\' e expansion of the parabolic cylinder
function, the second term of the finite sum Eq.(\ref{remrem}) and
the relation added to the leading part for possibility to perform an
infinite summation over the parabolic cylinder function:
\begin {eqnarray}
&&R(\mathcal{J}, K_0)=-\sum\limits_{j=0}^{\mathcal{J}}\;
\frac{(-1)^j }{j!\;(2z^2)^j}\ \sum_{k_i=K_0}^{\infty}\
\frac{\left(\xi\right)^{2k_{i}}}{(2k_{i})!}
\Gamma(k_{i-1}+k_{i}+1/2) \Gamma(k_{i}+k_{i+1}+2j+1/2)
\mathcal{D}_{-k_{i-1}-k_{i}-1/2}\;(z)
  +\nonumber \\
 &+&\sum_{k_i=0}^{K_0}\ \frac{\left(\xi\right)^{2k_{i}}}{(2k_{i})!}
\Gamma(k_{i-1}+k_{i}+1/2) \Gamma(k_{i}+k_{i+1}+1/2)
\mathcal{D}_{-k_{i-1}-k_{i}-1/2}\;(z) \
\epsilon_{\mathcal{J}}(k_{i}+k_{i+1},z) + \\
&+&\sum_{k_i=K_0+1}^{\infty}\
\frac{\left(\xi\right)^{2k_{i}}}{(2k_{i})!}
\Gamma(k_{i-1}+k_{i}+1/2) \Gamma(k_{i}+k_{i+1}+1/2)
\mathcal{D}_{-k_{i-1}-k_{i}-1/2}\;(z)\mathcal{D}_{-k_{i}-k_{i+1}-1/2}\;(z).\nonumber
\end {eqnarray}

The finite sum in remainder:
\begin {equation}
\sum_{k_i=0}^{K_0}\ \frac{\left(\xi\right)^{2k_{i}}}{(2k_{i})!}
\Gamma(k_{i-1}+k_{i}+1/2) \Gamma(k_{i}+k_{i+1}+1/2)
\mathcal{D}_{-k_{i-1}-k_{i}-1/2}\;(z) \
\epsilon_{\mathcal{J}}(k_i+k_{i+1},z)
\end {equation}
is bounded by the relation (\ref{temrem}):
\begin {equation}
\frac{\mathcal{M}}{(\mathcal{J}-1)!\
(2z^2)^{\mathcal{J}}}\sum_{k_i=0}^{K_0}\
\frac{\left(\xi\right)^{2k_{i}}}{(2k_{i})!}
\Gamma(k_{i-1}+k_{i}+1/2) \Gamma(k_{i}+k_{i+1}+2\mathcal{J}+1/2)
\mathcal{D}_{-k_{i-1}-k_{i}-1/2}\;(z) \
\end {equation}
where
\begin {equation}
\mathcal{M}=\max{\left(\frac{2z^2}{z^2-2a} \cdot_1F_2(\mathcal{J}/2,
1/2; \mathcal{J}/2+1;
1-\frac{a^2}{z^2})\exp{\left(\frac{4\delta}{z^2-2a}\
_1F_2(1/2,1/2;3/2;1-\frac{a^2}{z^2})\right)}\right)\ ,}
\end {equation}
and $a=k_i+k_{i+1}$, $k_{i\pm1}=1, 2, \dots, K_0\ .$ Let us remember
that $z^2\sim N^3.$ Since we have the freedom to choose the
parameter $\mathcal{J},$ the upper bound on this contribution to the
remainder can be made as small as necessary power of $N$.

In the asymptotic region of $ k_i>K_0$ we expand one of the function
$\mathcal{D}$ to double asymptotic expansions proposed by Temme
\cite{temme}:
\begin {equation}
\mathcal{D}_{-a-1/2}(z)=\frac{\exp{(-\mathcal{A}z^2)}}{(1+4\lambda)^{1/4}}
\left[\sum_{k=0}^{n-1}\frac{f_k(\lambda)}{z^{2k}}\ +\
\frac{1}{z^{2n}}R_n(a,z)\right]
\end {equation}
where the following quantities were introduced:
 $$\lambda = \frac{a}{z^2}\ ,\
 w_0 = \frac{1}{2}\; [\sqrt{1+4\lambda}-1]\ ,\
 \mathcal{A} =\frac{1}{2}\; w_0^2 + w_0 - \lambda - \lambda
 \ln{w_0} + \lambda\ln {\lambda}\ ,$$
 the functions $f_k(\lambda)$ are calculated in \cite{temme}.
We find for the infinite part of the sum  decompositions:

\begin {eqnarray}
&&\sum_{k_i=K_0+1}^{\infty}\
\frac{\left(\xi\right)^{2k_{i}}}{(2k_{i})!}
\Gamma(k_{i-1}+k_{i}+1/2) \Gamma(k_{i}+k_{i+1}+1/2)
\mathcal{D}_{-k_{i-1}-k_{i}-1/2}\;(z)\mathcal{D}_{-k_{i}-k_{i+1}-1/2}\;(z)\\
&=&\sum_{k_i=K_0+1}^{\infty}\
\frac{\left(\xi\right)^{2k_{i}}}{(2k_{i})!}
\Gamma(k_{i-1}+k_{i}+1/2) \Gamma(k_{i}+k_{i+1}+1/2)
\mathcal{D}_{-k_{i-1}-k_{i}-1/2}\;(z)\frac{\exp{(-\mathcal{A}z^2)}}{(1+4\lambda)^{1/4}}
\left[\sum_{k=0}^{n-1}\frac{f_k(\lambda)}{z^{2k}}\right]+\nonumber\\
&+&\sum_{k_i=K_0+1}^{\infty}\
\frac{\left(\xi\right)^{2k_{i}}}{(2k_{i})!}
\Gamma(k_{i-1}+k_{i}+1/2) \Gamma(k_{i}+k_{i+1}+1/2)
\mathcal{D}_{-k_{i-1}-k_{i}-1/2}\;(z)\frac{\exp{(-\mathcal{A}z^2)}}{(1+4\lambda)^{1/4}}
\left[ \frac{1}{z^{2n}}R_n(a,z)\right] \nonumber
\end {eqnarray}
It was shown in \cite{paper1} that the last part of this
contribution can be made as small as we need due to the freedom in
the choice of the parameter $n$ representing the number of the
functions $f_k(\lambda)$ taken into account in the Temme  double
asymptotic decomposition \cite{temme} of the parabolic cylinder
function.

Now we are going to estimate the last part of the remainder,
corresponding to the difference of the series:
\begin {eqnarray}
&&\sum_{k_i=K_0+1}^{\infty}\
\frac{\left(\xi\right)^{2k_{i}}}{(2k_{i})!}
\Gamma(k_{i-1}+k_{i}+1/2) \Gamma(k_{i}+k_{i+1}+1/2)
\mathcal{D}_{-k_{i-1}-k_{i}-1/2}\;(z)\frac{\exp{(-\mathcal{A}z^2)}}{(1+4\lambda)^{1/4}}
\left[
\sum_{k=0}^{n-1}\frac{f_k(\lambda)}{z^{2k}}\right]-\nonumber\\
&-&\sum\limits_{j=0}^{\mathcal{J}}\; \frac{(-1)^j }{j!\;(2z^2)^j}\
\sum_{k_i=K_0+1}^{\infty}\
\frac{\left(\xi\right)^{2k_{i}}}{(2k_{i})!}
\Gamma(k_{i-1}+k_{i}+1/2) \Gamma(k_{i}+k_{i+1}+2j+1/2)
\mathcal{D}_{-k_{i-1}-k_{i}-1/2}\;(z) \label{rem33}
\end {eqnarray}
Since both series are uniformly convergent, we exchange the order of
summations, to obtain:

\begin {eqnarray}
&&\sum_{k_i=K_0+1}^{\infty}\
\frac{\left(\xi\right)^{2k_{i}}}{(2k_{i})!}
\Gamma(k_{i-1}+k_{i}+1/2) \Gamma(k_{i}+k_{i+1}+1/2)
\mathcal{D}_{-k_{i-1}-k_{i}-1/2}\;(z)\nonumber\\
&&\left\{\frac{\exp{(-\mathcal{A}z^2)}}{(1+4\lambda)^{1/4}} \left[
\sum_{k=0}^{n-1}\frac{f_k(\lambda)}{z^{2k}}\right]-
\sum\limits_{j=0}^{\mathcal{J}}\;
\frac{(-1)^j\pochh{k_{i}+k_{i+1}+1/2}{2j} }{j!\;(2z^2)^j}\right\}\
 \label{rem4}
\end {eqnarray}
The double asymptotic expansion reduces to Poincar\' e - type
expansion \cite {temme2} when $a$ is fixed after expanding the
quantities in  $\lambda = a/z^2$ for small values of this parameter.
In the difference
\begin {equation}
\left\{\frac{\exp{(-\mathcal{A}z^2)}}{(1+4\lambda)^{1/4}} \left[
\sum_{k=0}^{n-1}\frac{f_k(\lambda)}{z^{2k}}\right]-
\sum\limits_{j=0}^{\mathcal{J}}\;
\frac{(-1)^j\pochh{k_{i}+k_{i+1}+1/2}{2j} }{j!\;(2z^2)^j}\right\}
\end {equation}
all terms where $k, j\leq \min{(n, \mathcal{J})}$ cancel one another
and the rest of terms is proportional or smaller a $z^{-2\min{(n,
\mathcal{J})}}$. As in the case of previous contributions to the
remainder, this means that this part of the remainder can be made as
small as we need in the power of $1/N$.

\section{Definition of the matrices}

The definition of the matrices $\ms A^d(\Lambda-1),\  \ms
C^d(\Lambda-1),\  \ms M^d(\Lambda-1)$ is the following:

1. The  $\ms A^d$ is the lower triangular matrix with  zeros over
the main diagonal of the dimension $(2\mu+1)(2\mu+1)$. The principal
minor of the dimension $(2d+1)(2d+1)$ is non-zero only with
elements:
$$\left\{\ms A^d(\Lambda-1)\right\}_{p,j}= \frac{a^{2d-j}_{2d-p}}{(A Q_{\Lambda}
Q_{\Lambda-1})^{p-j}}\ .$$

2. The $\ms C_{j,\lambda}^d(\Lambda) $ is the upper triangular
matrix with  zeros under the main diagonal of the dimension
$(2\mu+1)(\mu+1)$. The nonzero elements form the main minor of the
dimension $(2d+1)(d+1)$ with $\lambda^{th}$ column:
$$\ms C_{p,\lambda}^d(\Lambda)=(A Q_{\Lambda} Q_{\Lambda-1})^{2\lambda-p}
\left(\Lambda\right)_{2\lambda-p}^{2\lambda},$$ where $p = 0, 1,
..., 2\lambda$ and $\lambda = 0, 1, ..., d\ $.

3. The $\ms M_{\lambda,k}^d$ is the upper triangular matrix with
zeros under the main diagonal of the dimension $(\mu+1)(\mu+1)$. The
nonzero elements form the main minor of the dimension $(d+1)(d+1)\
$:
$$\ms M_{\lambda,k}^d(\Lambda-1) =
\binom{k}{\lambda}(Q^4_{\Lambda-1})^{k-\lambda} ,\  d\geq k\geq
\lambda \ge 0.$$

For a product of two lower-triangular matrices $\ms A^{I_3}(2)$ and
$\ms A^{I_2}(1)$ we have\cite{paper2}:

\begin {eqnarray}
\hspace{-2.0cm} &&\sum^p_{j=\lambda}\left\{\ms
A^{I_3}(2)\right\}_{p,j}\left\{\ms A^{I_2}(1)\right\}_{j,\lambda}=
\sum^p_{j=\lambda} \frac{a^{2I_3-j}_{2I_3-p}}{(AQ_3Q_2)^{p-j}}\
\frac{a^{2I_2-\lambda}_{2I_2-j}}{(AQ_2Q_1)^{j-\lambda}} =\\
\noalign{\bigskip}\nonumber
 &=&2^{-2(p-\lambda)} \frac{(4
I_2-2\lambda)!}{(4I_3-2p)!(p-\lambda)!}
\partial_\epsilon^{4i_2}\left\{(\epsilon^{4I_3-2p})
\sum^p_{j=\lambda}\binom{p-\lambda}{j-\lambda}\left(\frac{\epsilon^2}{AQ_3Q_2}\right)^{p-j}
\left(\frac{1}{AQ_2Q_1}\right)^{j-\lambda}\right\}_{\epsilon=1}\ ,
\end {eqnarray}
where $\epsilon$ is an auxiliary variable.

We have used  two identities for the summation over the index $j$
and the definition for index $I_j$ :

$$a^{2I_3-j}_{2I_3-p}\ a^{2I_2-\lambda}_{2I_2-j} =2^{-2(p-\lambda)}\
\frac{(4
I_2-2\lambda)!}{(4I_3-2p)!(p-\lambda)!}\binom{p-\lambda}{j-\lambda}
\frac{(4I_3-2j)!}{(4I_2-2j)!}$$ and
$$\frac{(4I_3-2j)!}{(4I_2-2j)!} = \partial_\epsilon^{4I_3-4I_2}(\epsilon^{4I_3-2j})|_{\epsilon=1}$$
 $$I_j = d - (i_j +i_{j+1} + \cdots +i_{\Lambda} )$$

In the above relation the summation over the index $j$ can be
performed explicitly. Introducing a new auxiliary variable:
$$\xi = \epsilon^2 ,$$

we find:

$$\sum^p_{j=\lambda}\left\{\ms A^{I_3}(2)\right\}_{p,j}\left\{\ms A^{I_2}(1)\right\}_{j,\lambda} =
2^{-2(p-\lambda)} \frac{(4 I_2-2\lambda)!}{(4I_3-2p)!(p-\lambda)!}
2^{4i_2}D^{i_2}_{\xi}\left\{\frac{1}{\xi^{p-2I_3}}
\left(\frac{\xi}{AQ_3Q_2}+
\frac{1}{AQ_2Q_1}\right)^{p-\lambda}\right\}_{\xi=1} \ ,$$

where $D_{\xi}=3/4\partial^2_{\xi}+3 \xi \partial^3_{\xi}+\xi^2 \partial^4_{\xi}$ is calculated from
$\partial^4_{\epsilon}$ .

For  the resulting product of all matrices $\ms A^I(k)$ we find:

$$\left\{\ms A^{d-i_{\Lambda}}(\Lambda-1)*\ms A^{d-i_{\Lambda}-i_{\Lambda-1}}(\Lambda-2)*
\cdots
*\ms A^{d-i_{\Lambda}-i_{\Lambda-1}-\cdots-i_2}(1)\right\}_{p,\lambda}=$$
$$
= 2^{-2(p-\lambda)}
\frac{(4I_2-2\lambda)!}{(4I_{\Lambda}-2p)!(p-\lambda)!}
2^{4(i_2+i_3+\cdots+i_{\Lambda})}\times$$

$$\times\left\{\prod_{m=2}^{\Lambda}D^{i_m}_{\xi_m}
\left[\frac{1}{\xi_m^{p-2I_{m+1}}} \left(
\frac{1}{AQ_2Q_1}+\frac{\xi_2}{AQ_3Q_2}+\cdots+
\frac{\xi_2\cdots\xi_{\Lambda-1}}{AQ_{\Lambda}Q_{\Lambda-1}}
\right)^{p-\lambda} \right]\right\}_{(all \xi_m \rightarrow
1)}
$$
By symbol $*$ we indicate the product of matrices.
Evaluating the product of matrices:

$$\left\{\tilde{\ms M}^{d-i_{\Lambda}-i_{\Lambda-1}-\cdots-i_2}(1)*\cdots*
\tilde{\ms M}^{d-i_{\Lambda}-i_{\Lambda-1}}(\Lambda-2)* \tilde{\ms
M}^{d-i_{\Lambda}}(\Lambda-1)\right\}$$

we use that $\tilde{\ms M}^{I_{j+1}}(j)$ is a one-column matrix with
$I_{j+1}$ non-zero elements in the $(j+1)$ column:

$$\left\{\tilde{\ms M}^{I_{j+1}}(j)\right\}_{\lambda,j+1} =
\binom{I_{j+1}}{\lambda}Q_j^{4(I_{j+1}-\lambda)},$$ where
$\lambda=0, 1, \cdots , I_{j+1}.$ Product of such matrices is a
one-column matrix with  elements:

$$\left\{\tilde{\ms M}^{d-i_{\Lambda}-i_{\Lambda-1}-\cdots-i_2}(1)*\cdots*
\tilde{\ms M}^{d-i_{\Lambda}-i_{\Lambda-1}}(\Lambda-2)* \tilde{\ms
M}^{d-i_{\Lambda}}(\Lambda-1)\right\}_{\lambda,I_{\Lambda}}=$$

$$=\binom{I_2}{\lambda}\binom{I_3}{I_2} \cdots \binom{I_{\Lambda}}{I_{\Lambda-1}}
Q_1^{4(I_2-\lambda)}Q_2^{4(I_3-I_2)}\cdots
Q_{\Lambda-1}^{4(I_{\Lambda}-I_{\Lambda-1})}$$

From the definition (\ref{recu}) of  recurrence steps we have for
the matrix $\ms C^{I_2}(1)$  nonzero elements:

$$\left\{\ms C^{I_2}(1)\right\}_{j,\lambda}=
\frac{Q_0^{4\lambda}}{(AQ_1Q_0)^j}a^{2\lambda}_{2\lambda-j}\ ,$$ with
 conditions for indexes: $$0\leq j \leq 2\lambda \leq 2I_2 \leq
2\mu\ .$$

 Collecting all partial results together, inserting them into Eq. (\ref{matr1}) and remembering that for
function $\mathcal{S}_{\Lambda}$ defined in Eq. (\ref{rov12}) only
matrix elements $\left\{\ms C(\Lambda)^{2\mu}\right\}_{2\mu, 2\mu}$
are important, we find the result (\ref{maineq}).


\begin{thebibliography}{99}

\bibitem{tusz}Tuszy\' nski J.A., Clouter M.J., and Kiefe H., \textit{Non -
Gaussian models for critical fluctuations}, Phys. Rev. \textbf{B35}
(1986) 3423.
\bibitem{prud}Prudnikov A.P., Britchkov J.A., Marichev O.I., \textit{Integrals and Series},
in Russian Nauka 1981, in English Gordon and Breach, New York 1986.
\bibitem{bateman} Bateman H., \textit{Higher Transcendental Functions}, Volume
II, Mc Graw-Hill, 1953.
\bibitem{dem} Chaichian M., Demichev A., \textit{Path Integrals in
Physics}, Vol. I, IOP Publishing Ltd. 2001.
\bibitem{paper1}  J. Boh\' a\v cik and  P. Pre\v snajder, Functional
integral for $\varphi^4$ potential beyond classical perturbative
methods, hep-th/0503235
\bibitem{bkz} Gelfand I.M. and Yaglom A.M., J. Math. Phys. 1
(1960) 48.
\bibitem{olver} Olver F.W.J., \textit{Uniform asymptotic expansions for Weber parabolic cylinder functions
of large order}, J. Research NBS, 63B:131-169, 1959.
\bibitem{temme2}Vidunas R., Temme N.M., \textit{Parabolic cylinder functions: Examples of error
bounds for asymptotics expansions}, report MAS-R0225 October 31,
2002.
\bibitem{paper2}  J. Boh\' a\v cik and  P. Pre\v snajder, Functional
integral for $\varphi^4$ potential beyond classical perturbative
methods II, arXiv:0711.4683.
\bibitem{wolf} Wolfram S., Mathematica, Addison-Wesley, 1991.
\bibitem{kamke} Dr. E. Kamke, Differentialgleichungen Losungsmethoden und Losungen, 6. verbesserte aulage, Leipzig 1959,
in Russian Nauka, Moscow 1965.
\bibitem{lewis} H.R.Lewis, \textit{Phys. Rev. Let.} \textbf{27}, 510
(1967).
\bibitem{samaj} L. \v Samaj, \textit{Int. Journal of Mod. Phys.}
\textbf{B16}, 3909-3914 (2002).
\bibitem{bender} C.M.Bender and T.T.Wu, \textit{Phys. Rev.}
\textbf{184}, 1231 (1969).
\bibitem{findif} Milne-Thomson L.M., \textit{The calculus of finite
differences},  First Edition 1933.
\bibitem{temme} Temme N.M., \textit{Numerical and Asymptotic Aspect of
Parabolic Cylinder Functions}, J. of Computational and Applied
Math., 121 (2000) 221-246.



\end{thebibliography}
\end{document}